\documentclass[letterpaper]{article} 
\usepackage{aaai24}  
\usepackage{times}  
\usepackage{helvet}  
\usepackage{courier}  
\usepackage[hyphens]{url}  
\usepackage{graphicx} 
\usepackage{natbib}  
\usepackage{caption} 
\usepackage{algorithm}
\usepackage{algorithmic}

\usepackage{newfloat}
\usepackage{listings}
\DeclareCaptionStyle{ruled}{labelfont=normalfont,labelsep=colon,strut=off} 
\floatstyle{ruled}
\newfloat{listing}{tb}{lst}{}
\floatname{listing}{Listing}
\usepackage{natbib}
\usepackage{breqn}
\usepackage{booktabs}
\usepackage{multirow}
\usepackage{makecell}
\usepackage{amsmath}
\usepackage[all]{nowidow}
\usepackage{subcaption}

\newcommand\blfootnote[1]{%
  \begingroup
  \renewcommand\thefootnote{}\footnote{#1}%
  \addtocounter{footnote}{-1}%
  \endgroup
}

\title{Dynamics of Moral Behavior in Heterogeneous Populations of Learning Agents}

\author{Elizaveta Tennant\textsuperscript{\rm 1},
    Stephen Hailes\textsuperscript{\rm 1},
    Mirco Musolesi\textsuperscript{\rm 1, \rm 2}
     }

\affiliations{
    \textsuperscript{\rm 1}Department of Computer Science\\
    University College London, Gower Street, London WC1E 6BT, United Kingdom \\
    \textsuperscript{\rm 2}Department of Computer Science and Engineering\\
    University of Bologna, Viale del Risorgimento 2, 40133 Bologna, Italy 
 
    l.karmannaya.16@ucl.ac.uk,
    s.hailes@ucl.ac.uk, 
    m.musolesi@ucl.ac.uk
}

\begin{document}

\maketitle

\begin{abstract}
Growing concerns about safety and alignment of AI systems highlight the importance of embedding \textit{moral} capabilities in artificial agents: a promising solution is the use of learning from experience, i.e., Reinforcement Learning. In multi-agent (social) environments, complex population-level phenomena may emerge from interactions between individual learning agents. Many of the existing studies rely on simulated social dilemma environments to study the interactions of independent learning agents; however, they tend to ignore the moral \textit{heterogeneity} that is likely to be present in societies of agents in practice. For example, at different points in time a single learning agent may face opponents who are consequentialist (i.e., focused on maximizing outcomes over time), norm-based (i.e., conforming to specific norms), or virtue-based (i.e., considering a combination of different virtues). The extent to which agents' co-development may be impacted by such moral heterogeneity in populations is not well understood. In this paper, we present a study of the learning dynamics of morally heterogeneous populations interacting in a social dilemma setting. Using an Iterated Prisoner's Dilemma environment with a partner selection mechanism, we investigate the extent to which the prevalence of diverse moral agents in populations affects individual agents' learning behaviors and emergent population-level outcomes. We observe several types of non-trivial interactions between pro-social and anti-social agents, and find that certain types of moral agents are able to steer selfish agents towards more cooperative behavior. 
\end{abstract}


\section{Introduction}
\label{sec:Introduction}

Growing concerns about the safety and alignment of Artificial Intelligence (AI) systems highlight the importance of embedding pro-social capabilities into artificial agents. Without these, agents might potentially behave harmfully in multi-agent situations such as social dilemmas (e.g., see \citealt{leib02017multiagent}). 
In particular, in this work, we focus on modeling the development of \textit{moral decision-making}. 

In general, it has been shown that morality can be developed in agents through learning from experience \citep{hughes2018inequity,mckee2020social,tennant2023modeling}. Learning approaches to moral alignment offer a variety of advantages, including potentially greater adaptability and generality \citep{tennant2024learning}, and the ability to learn implicit preferences \citep{ouyang2022training}. Moral principles can range from consequentialist \citep{Bentham1996} to norm-based morality \citep{kant1981grounding}, and from entirely pro-social to entirely anti-social preferences. Furthermore, the morality of agents can be based on a combination of virtues, as exemplified by the Virtue Ethics approach \citep{aristotle}. Existing works have proposed a variety of individual models to predispose agents to specific moral principles (e.g., \citealt{Bazzan1999moralsentiments, Capraro2021}), and recent studies have demonstrated that these can be effectively encoded in Reinforcement Learning (RL) agents via intrinsic rewards \citep{hughes2018inequity,mckee2020social,tennant2023modeling}. 
\blfootnote{Version with Appendix: \small{\url{https://arxiv.org/abs/2403.04202}}}
\blfootnote{Code: \small{\url{https://github.com/Liza-Tennant/moral_MARL}}}

However, to date, we still have limited understanding of how \textit{different} types of morality may co-evolve in heterogeneous populations. 
In fact, many real-world AI systems are likely to co-exist (essentially forming systems of systems) and may be co-developed in parallel with others. Especially when it comes to integrating moral decision-making into systems, different stakeholders may decide to prioritize varying principles or preferences. It is therefore crucial that we start developing an understanding of how the presence of different moral preferences in populations affects individual agents' learning behaviors and interactions, and, indeed, emergent population-level outcomes. Furthermore, simulation studies such as ours contribute to raising awareness about emergent behaviors and the possibility of unintuitive outcomes emerging from multi-agent learning scenarios.

This type of agent-based analysis of moral learning may additionally offer insights into the potential dynamics of interactions in human societies in the tradition of computational philosophy \citep{alexander2007structural}, provide an experimental test-bed for new theories \citep{MayoWilsonConor2021Tcps}, or simulate potential evolutionary mechanisms underpinning human moral preferences, in a similar way to Evolutionary Game Theory \citep{hofbauer_sigmund_1998,sigmund1999evolutionary,sigmund2010}. 

The core contribution of this work is the \textit{study of behavior and population dynamics among RL agents with diverse moral preferences}, providing insights for the design of artificial agents with a focus on safety and alignment. More generally, we present a \textit{methodology for analyzing emergent behavior in populations of agents with heterogeneous moral orientation}.
We apply the proposed methodology to the Iterated Prisoner's Dilemma (IPD, \citealt {rapoport1974prisoner}), demonstrating at the same time its potential applicability and generalizability to a variety of scenarios and different moral frameworks.


\section{Background and Preliminaries}
\label{sec:Background}



\subsection{Social Dilemma Games}

Social dilemma games simulate social situations in which players obtain different utilities (payoffs) from choosing one action or another, and the structure of these utilities is such that each player faces a trade-off between individual interest and societal benefit when choosing an action \citep{dawes1980}. The most widely studied type is a symmetric matrix game with two players and two abstract actions - Cooperate (C) or Defect (D). Players in these games must decide on their respective actions simultaneously, without communicating. 

A classic game from Economics and Philosophy, which is relevant to moral choice, is the Prisoner's Dilemma (\citealt{rapoport1974prisoner}, see payoffs for row vs column player in Table \ref{tab:game}). 
We implement the \textit{Iterated} Prisoner's Dilemma (IPD)
in a population, in which agents interact in pairs in discrete time steps and aim to maximize their cumulative payoff over time. In the iterated version of the game, players can take actions to punish their opponents for past defection or to influence their future behaviors. 

\begin{table}[h]
\begin{center}
\begin{tabular}{l|cc}
\textbf{IPD} & C    & D    \\ \hline
C         & 3, 3 & 0, 4 \\ 
D         & 4, 0 & 1, 1 
\end{tabular}
\end{center}
\caption{Payoffs on one step of the \textit{Iterated Prisoner's Dilemma (IPD)} game for the row \& column player.}
\label{tab:game}
\end{table}

Deriving predicted equilibria for these games is not always feasible: these are systems composed of interacting (heterogeneous) entities, where learning dynamics may cause instability in the environment and dynamic behaviors \citep{busoniubabuska2008}. Therefore, simulation is needed to study potential emergent behaviors and outcomes \citep{Anderson1972more}.



\subsection{Deep Reinforcement Learning in Markov Games}

Reinforcement Learning (RL) is a well-suited technique for modeling agents that learn by interacting with others in an environment \citep{Sutton2018RLSE}. It can be applied in conjunction with Evolutionary Game Theory \citep{hofbauer_sigmund_1998} to iterated social dilemma games \citep{littman1994markov,sandholm1996multiagent,de2006learning, abel2016reinforcement}, in which game payoffs constitute \textit{extrinsic} rewards, and traits such as moral or social preferences \citep{fehr2002social} can be encoded in the agent's \textit{intrinsic} reward \citep{chentanez2004intrinsically}. In the following section, we discuss the design of intrinsic moral rewards in detail. It is also worth noting that we assume populations of independent, continuously learning agents. This creates interesting dynamics as agents affect one another's learning process \citep{leibo2019autocurricula}. Given the potentially large number of states, we adopt DQN as the underlying learning algorithm \citep{mnih2015human}. 

\subsection{Morality as Intrinsic Reward}

Traditional social dilemma scenarios assume that agents are only motivated by accumulating game payoffs (i.e., extrinsic reward). However, human data has shown that other preferences may also come into play, such as the predisposition to cooperation \citep{camerer2011behavioral}. In artificial agents, preferences other than game rewards can be encoded in intrinsic rewards \citep{chentanez2004intrinsically}. As a result, recent multi-agent RL studies have been focused on modeling various pro-social preferences as intrinsic reward functions for social dilemma players \citep{hughes2018inequity, peysakhovich2018consequentialist, mckee2020social}. Moral rewards in particular have been studied for two-agent (i.e., dyadic) interactions in \citet{tennant2023modeling}.

We anchor our intrinsic reward definitions in traditional moral philosophical frameworks, especially relying on the distinction between \textit{consequentialist} versus \textit{norm-based} morality, and the idea of \textit{virtue ethics}. \textit{Consequentialist} morality focuses on the consequences of an action, and includes Utilitarianism \citep{Bentham1996}, which defines actions as moral if they maximize total utility for all agents in a society. \textit{Norm-}based morality, including Deontological ethics \citep{kant1981grounding}, considers an act moral if it does not contradict the society's external norms. Finally, in \textit{Virtue Ethics} \citep{aristotle}, moral agents must act in line with their certain internal virtues, such as fairness or care for others \citep{graham2013moral}. Different virtues can matter more or less to different agents \citep{aristotle,graham2009liberals} and can themselves have consequentialist or norm-based foundations. We present the specific set of agents considered in this study and their classification in the Methodology section. 


\subsection{Partner Selection}
In human populations, agents have a choice of which individual to interact with. 
Given this selection mechanism, reputation comes into play and competitive and collaborative relationships may form. \citet{Santos2008} show experimentally that these dynamics are likely to lead to a re-structuring of the population. Adaptive behavior resulting from selection mechanisms has been hypothesized to drive the emergence of cooperation \citep{BarclayWiller2007partnerchoice, Cuesta2015reputation}. In the context of societies of artificial learning agents, these mechanisms have recently been studied by \citet{Anastassacos2020partner} and \citet{baker2020team}. 

In particular, when applying partner selection to the IPD, a conflict may arise between a player's motivation to appear cooperative in order to get selected more often by other payers, and the motivation to select and then exploit cooperators to gain a greater payoff. \citet{Anastassacos2020partner} find partner selection to be norm-inducing and even lead to the emergence of cooperation in a population of purely selfish agents. We adopt a partner selection model similar to \citet{Anastassacos2020partner}.

Selection dynamics in morally heterogeneous populations are likely to be more complex and harder to predict. Since each player plays according to their own intrinsic reward signal, different coalitions \citep{Shenoy1979OnCF} may arise among different agent types within a population, and popularity of agents may not directly correlate with cooperativeness. Running simulation experiments to study this may provide insight into the types of behaviors and outcomes that may develop in heterogeneous populations with selection, with implications not only for the study of artificial agents, but also of human and animal societies and Evolutionary Game Theory \citep{sigmund1999evolutionary}.

\subsection{Learning in Heterogeneous Populations}

Evidence from the social sciences suggests that human societies are morally heterogeneous \citep{graham2013moral,bentahila2021universality}. While certain commonsense norms may be agreed upon by a society \citep{Reid1990,Gert2004}, distinct moral principles are likely to hold a different weight for different individuals depending on factors such as political orientation \citep{graham2009liberals}, personality \citep{Lifton1985personality}, nurture or even natural predispositions \citep{sinnott2008moral}. 
%
%
%
%
%

\citet{mckee2020social} conducted a social dilemma study with diverse consequentialist agents (i.e., agents with various preferences over group outcome distributions). Their findings suggest that agents trained in heterogeneous populations develop particularly generalized and high-performing policies. In contrast, populations containing only altruistic agents are characterized by greater collective reward, but at the cost of equality between agents. In addition, in \citet{McKee2022quantifying}, the authors define population diversity as the set of various opponents' policies that an agent may face, and also find that training in diverse populations can lead to improvements in terms of agent performance on some types of environments. 

Even without learning, \citet{Santos2008} show that - in fixed networks of agents - diversity in terms of neighborhoods within the population graph promotes cooperation in the population. 
With respect to global societal outcomes, \citet{Ord2015MoralTrade} discusses how, through interaction, morally diverse agents taking actions to satisfy their distinct moral goals may improve the global welfare of the whole population, akin to a `trade'.
We believe that these conceptual frameworks can represent the basis for studying how artificial agents may learn to act across populations containing different proportions of opponents with different moral preferences. 




\section{Methodology}
\label{sec:Methodology}

\subsection{Experimental Setup}

Our environment involves a population $P$ (of size $N=16$) playing an iterated game. At each iteration, a moral player $M$ and an opponent $O$ play a simultaneous one-shot Prisoner's Dilemma game with two possible actions - Cooperate or Defect ($a^t_M, a^t_O \in \{C, D\}$, see Table \ref{tab:game} for the respective payoffs). To allow learning in a population, we run our simulation in 30000 episodes, where a single episode involves each player selecting a partner once, and then each corresponding pair of players playing the game. For clarity, each agent gets to make a selection on each episode, so $N$ games are played. In particular, at iteration $t$, the learning agent $M$ observes a selection state including the action played by each possible opponent $O \in P$ at $t-1$: $s_{M,sel}^{t}=[a_{O_1}^{t-1}, ..., a_{O_{n-1}}^{t-1}]$. Using this state and the current learned policy, $M$ selects an opponent $O$. 
Then, to choose an action for the dilemma, player $M$ relies on the state representing the latest move of their selected opponent $O$: $s_{M,dil}^{t}=a_O^{t-1}$. Simultaneously, their opponent $O$ chooses an action following the same principle. The players $M$ and $O$ then each receive a game reward $R_M^{t+1}$ and $R_O^{t+1}$ (corresponding to the game payoffs) and observe a new state $s_{M,dil}^{t+1}$ and $s_{O,dil}^{t+1}$ based on their opponent's move $a^t$.

In this Markov game \citep{littman1994markov}, over time both players learn to make selections and take actions by observing past interactions with various opponents. 
Each agent uses Q-Learning \citep{watkins1992q} to update the state-action value function $Q$ (i.e., an estimate of the cumulative expected return over time). Given a large number of possible selection states, we approximate $Q$ using a neural network. Each agent maintains two internal models (one for partner selection, and one for playing the dilemma) - for consistency, we use function approximation for learning both models, and each is parameterized by a fully connected network with a single hidden layer of size $256$.

\begin{table*}[t]
  \small
  \centering
  \begin{tabular}{lllll}\toprule %
    & \multicolumn{2}{l}{\textit{Agent Label \& Moral Type}} & \textit{Moral Reward Function} & \textit{Source of Morality} \\ \toprule  
    & \\[-1.5ex]
    & \textit{S} & Selfish & None (use $R_{M_{extr}}^t$ to learn) & None \\[-1.5ex]
    \\
    \hline  & \\[-1.5ex]
    & \textit{Ut} & Utilitarian  & $R_{M_{intr}}^t=R_{M_{extr}}^t + R_{O_{extr}}^t$ & External/Internal consequentialist\\ 
    \multirow{3}{*}{\makecell[ll]{\rotatebox[origin=c]{90}{\textit{Pro-Social}}}}
    & \textit{De} & Deontological  & $R_{M_{intr}}^t= 
    \begin{cases}
        $--$\xi,& \text{if } a_M^t=D ,  a_O^{t-1}=C \\ 
        0,              & \text{otherwise}
    \end{cases}\ $ & External norm \\ 
    & \textit{V-Eq} & Virtue-Equality & $R_{M_{intr}}^t=1-\frac{|R_{M_{extr}}^t-R_{O_{extr}}^t|}{R_{M_{extr}}^t+R_{O_{extr}}^t}$ & Internal consequentialist \\ 
    & \textit{V-Ki} & Virtue-Kindness  & $R_{M_{intr}}^t= 
    \begin{cases}
        \xi,& \text{if } a_M^t=C  \\
        0,              & \text{otherwise}
    \end{cases}\ $ & Internal norm  \\ 

    \hline & \\[-1.5ex]
    & \textit{aUt} & Anti-Utilitarian  & $R_{M_{intr}}^t=-(R_{M_{extr}}^t + R_{O_{extr}}^t)$ & External/Internal consequentialist\\ 
    \multirow{3}{*}{\makecell[ll]{\rotatebox[origin=c]{90}{\textit{Anti-Social}}}}
    & \textit{mDe} & Malicious Deontological  & $R_{M_{intr}}^t= 
    \begin{cases}
        \xi,& \text{if } a_M^t=D ,  a_O^{t-1}=C \\ 
        0,              & \text{otherwise}
    \end{cases}\ $ & External norm \\
    & \textit{V-In} & Virtue-Inequality & $R_{M_{intr}}^t=\frac{|R_{M_{extr}}^t-R_{O_{extr}}^t|}{R_{M_{extr}}^t+R_{O_{extr}}^t}$ & Internal consequentialist \\ 
    & \textit{V-Ag} & Virtue-Aggression  & $R_{M_{intr}}^t= 
    \begin{cases}
        \xi,& \text{if } a_M^t=D  \\
        0,              & \text{otherwise}
    \end{cases}\ $ & Internal norm  \\ 

    \bottomrule
  \end{tabular}
  \caption{Definitions of the types of intrinsic moral rewards, from the point of view of the moral agent $M$ playing versus an opponent $O$ at iteration $t$. We define four pro-social players, four anti-social ones, and the traditional \textit{Selfish} player.}
  \label{tab:moralpayers}
\end{table*}

For learning both selections and the dilemma, players record the experiences $(s,a,r,s')$ in their memory buffer $D_{sel}$ and $D_{dil}$. We copy the reward obtained from the game into both the dilemma and selection memories. At the end of an episode, all players simultaneously update the Q-network parameters $\theta_t$ for their two models using the latest experience available in the memory buffer $D$, and the Mean Squared Error loss function \citep{mnih2015human}, where $\gamma = 0.99$ is a discount factor: 
\begin{multline}
L_t(\theta_t) = {E}_{s,a,r,s'} \bigg [\bigg(R^{t+1}+\gamma\max_a Q(s^{t+1},a, \theta_t) \\ -Q(s^t,a^t,\theta_t)\bigg)^2 \bigg]
\end{multline}

If more than one experience is available (i.e., a player plays more than one dilemma game in that episode), we calculate an average loss across the experiences before updating the network. Each memory buffer $D$ is refreshed before the start of the next episode, so that each player only learns from the latest episode of experience. In updating the network weights with the Adam optimizer \citep{kingma2017adam}, we use a learning rate of $0.001$. 


Agents select partners and play the dilemma using an $\epsilon$-greedy policy, acting randomly with probability $\epsilon$ or otherwise playing greedily according to the Q-values learned so far: 
\begin{eqnarray}
    \pi(s_t)= 
\begin{cases}
    \arg\max_a Q(s_t,a) , & \text{with probability } 1-\epsilon  \\ 
    U(A=\{C,D\}),         & \text{with probability } \epsilon .
\end{cases}\
\label{eq:policy}
\end{eqnarray}
We set $\epsilon_{sel}=0.1$ and $\epsilon_{dil}=0.05$. The full detailed algorithm can be found in the Appendix.

Player $M$ can learn according to an \textit{extrinsic} game reward $R_{M_{extr}}^{t+1}$, which depends directly on the joint actions $a_M^t,a_O^t$ (as defined in Table \ref{tab:game}), or according to an \textit{intrinsic} moral reward $R_{M_{intr}}^{t+1}$.
In the next subsection, we discuss the definition of these intrinsic moral rewards.

\subsection{Modeling Morality as Intrinsic Reward}
\label{subsec:moralrewards}

Intrinsic rewards associated with pro-social preferences have been used to incentivize the emergence of cooperation in social dilemmas \citep{hughes2018inequity, peysakhovich2018consequentialist, jaques2019social}. \citet{tennant2023modeling} extended this work by defining four Q-learning moral\footnote{We use the term moral for agents that are not selfish. However, selfishness itself can be considered as a moral choice, expressed as rational egotism. In the case of social dilemmas, selfishness maps to the concept of rationality. For these agents $R_{M_{intr}} = R_{M_{extr}}$.} agents that learn according to various intrinsic rewards $R_{M_{intr}}$, and contrasting these against a traditional \textit{Selfish} agent which learns to maximize its extrinsic (game) reward $R_{M_{extr}}$ (e.g., \citealt{leib02017multiagent}).

In this study, we rely on eight types of moral agents - four pro-social moral learners from \citet{tennant2023modeling} \textit{(Utilitarian, Deontological, Virtue-Equality, Virtue-Kindness)}, and four additional anti-social counterparts \textit{(anti-Utilitarian, malicious-Deontological, Virtue-Inequality, Virtue-Aggression)}. The agents are defined as follows (for formal definitions of the reward functions, please refer to Table \ref{tab:moralpayers}):
\begin{itemize}
\item the \textit{Utilitarian (Ut)} agent tries to maximize the collective payoff (i.e., total payoff for both players; \citealt{Bentham1996});
\item the \textit{Deontological (De)} agent tries to follow the norm of conditional cooperation \citep{kant1981grounding,fehr2004social} and gets penalized through the negative reward $-\xi$ for defecting against a cooperator (i.e., an opponent who previously cooperated); 
\item the \textit{Virtue-Equality (V-Eq)} agent tries to maximize equality between the two players' payoffs $R_{M_{extr}}^t$ and $R_{O_{extr}}^t$, measured using a two-agent variation of the Gini coefficient \citep{gini1912variabilita};
\item the \textit{Virtue-Kindness (V-Ki)} agent receives a reward $\xi$ for acting kindly (i.e., cooperating) against any opponent \citep{aristotle};
\item the \textit{anti-Utilitarian (aUt)} agent tries to minimize the collective payoff;
\item the \textit{malicious-Deontological (mDe)} agent tries to follow the conditional defection norm and receives a reward $\xi$ for defecting against an cooperator; 
\item the \textit{Virtue-Inequality (V-In)} agent tries to minimize equality between the two players' payoffs $R_{M_{extr}}^t$ and $R_{O_{extr}}^t$;
\item the \textit{Virtue-Aggression (V-Ag)} agent receives a reward $\xi$ for acting aggressively (i.e., defecting) against any opponent.
\end{itemize}

Agent types \textit{Ut} and \textit{aUt} can be defined as external consequentialist since their reward depends on consequences in the environment (i.e., the players' rewards). The \textit{De} and \textit{mDe} agents depend on an external reputation-based norm defined in terms of current actions. \textit{V-Eq} and \textit{V-In} agents can be considered internal consequentialist, since they follow a consequence-based internal virtue. Finally, \textit{V-Ki} and \textit{V-Ag}'s pre-dispositions originate from the agent's internal norm.

\begin{table*}[h]
  \small
  \centering
  \begin{tabular}{ll}\toprule
    \textit{Population Label} & \textit{Population Composition} \\ \toprule
     \textit{majority-S} & 8x\textit{S}, 1x\textit{Ut}, 1x\textit{aUt}, 1x\textit{De}, 1x\textit{mDe}, 1x\textit{V-Eq}, 1x\textit{V-In}, 1x\textit{V-Ki}, 1x\textit{V-Ag} \\
     \textit{majority-Ut}  & 1x\textit{S}, 8x\textit{Ut}, 1x\textit{aUt}, 1x\textit{De}, 1x\textit{mDe}, 1x\textit{V-Eq}, 1x\textit{V-In}, 1x\textit{V-Ki}, 1x\textit{V-Ag} \\ 
     \textit{majority-aUt} & 1x\textit{S}, 1x\textit{Ut}, 8x\textit{aUt}, 1x\textit{De}, 1x\textit{mDe}, 1x\textit{V-Eq}, 1x\textit{V-In}, 1x\textit{V-Ki}, 1x\textit{V-Ag} \\ 
     \textit{majority-De} & 1x\textit{S}, 1x\textit{Ut}, 1x\textit{aUt}, 8x\textit{De}, 1x\textit{mDe}, 1x\textit{V-Eq}, 1x\textit{V-In}, 1x\textit{V-Ki}, 1x\textit{V-Ag} \\ 
     \textit{majority-mDe} & 1x\textit{S}, 1x\textit{Ut}, 1x\textit{aUt}, 1x\textit{De}, 8x\textit{mDe}, 1x\textit{V-Eq}, 1x\textit{V-In}, 1x\textit{V-Ki}, 1x\textit{V-Ag} \\ 
     \textit{majority-V-Eq} & 1x\textit{S}, 1x\textit{Ut}, 1x\textit{aUt}, 1x\textit{De}, 1x\textit{mDe}, 8x\textit{V-Eq}, 1x\textit{V-In}, 1x\textit{V-Ki}, 1x\textit{V-Ag} \\ 
     \textit{majority-V-In} & 1x\textit{S}, 1x\textit{Ut}, 1x\textit{aUt}, 1x\textit{De}, 1x\textit{mDe}, 1x\textit{V-Eq}, 8x\textit{V-In}, 1x\textit{V-Ki}, 1x\textit{V-Ag} \\ 
     \textit{majority-V-Ki} & 1x\textit{S}, 1x\textit{Ut}, 1x\textit{aUt}, 1x\textit{De}, 1x\textit{mDe}, 1x\textit{V-Eq}, 1x\textit{V-In}, 8x\textit{V-Ki}, 1x\textit{V-Ag} \\ 
     \textit{majority-V-Ag} & 1x\textit{S}, 1x\textit{Ut}, 1x\textit{aUt}, 1x\textit{De}, 1x\textit{mDe}, 1x\textit{V-Eq}, 1x\textit{V-In}, 1x\textit{V-Ki}, 8x\textit{V-Ag} \\ 
    \bottomrule
  \end{tabular}
  \caption{Populations considered in this study. Each population contains eight players of a certain `majority' type and one player of each other type.}
  \label{tab:populations}
\end{table*}

The rewards in Table \ref{tab:moralpayers} are defined for a single iteration $t$. Given the payoff matrices of the IPD game used (see Table \ref{tab:game}), we set the parameter $\xi$ in the four norm-based rewards to be $\xi=5$, so it sends a strong signal of a value similar to the maximum game payoff available. 
Through further experiments, we also observe that the choice of smaller values of $\xi$ does not affect the overall results.

\subsection{Population Compositions}

We compare a set of heterogeneous populations based on the following principles: population size is equal to 16; all populations contain at least one player of each type; one player type constitutes the majority (i.e., 8 out of 16 players). Given 9 possible player types, this results in 9 possible population compositions, as outlined in Table \ref{tab:populations}. 


\section{Results}
\label{sec:Results}



We separately analyze cooperation (at population and individual level), social outcomes (at population level), selection dynamics and rewards obtained by different player types across the nine populations.

\subsection{Emergence of Cooperation}\label{subsec:Population}


\begin{figure}[h] 
    \centering
        (a) \includegraphics[height=0.58\linewidth]{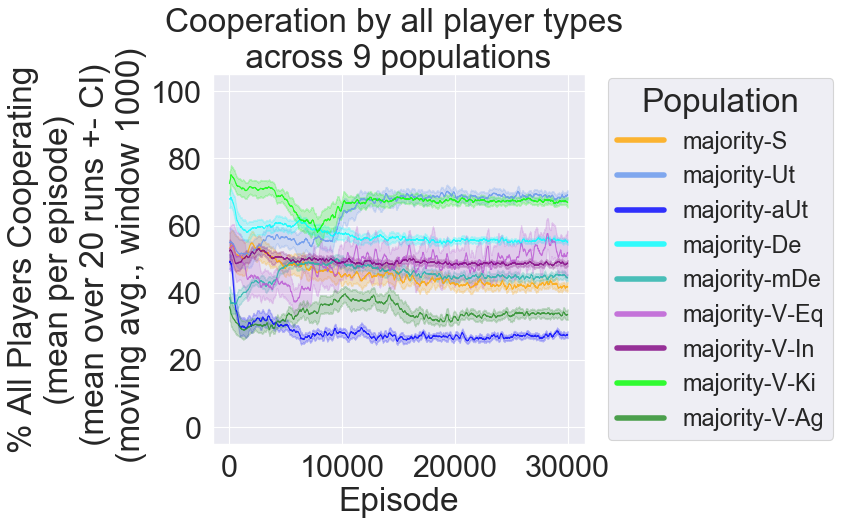}
        \\
        (b) \includegraphics[height=0.58\linewidth]{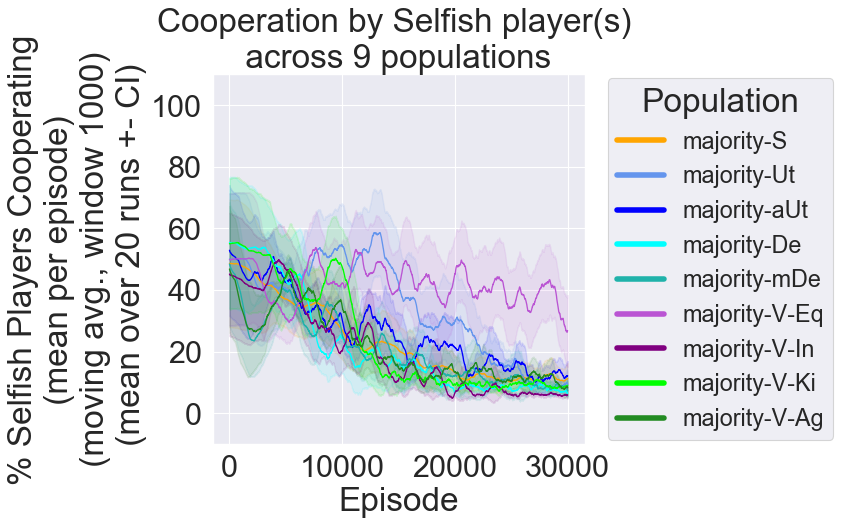}
    
    \caption{\textbf{(a)} Cooperation by all player types within each population over time. \textbf{(b)} Cooperation by the \textit{Selfish} player(s) in every population over time. In these charts, we plot the moving average of the mean across 20 runs.}
    \label{fig:Cooperation}
\end{figure}

\subsubsection{Cooperation across the entire population.}

We first study whether a stronger prevalence of certain agent types in every population leads to higher levels of cooperation in the population as a whole. Cooperation over time for each population is presented in Figure \ref{fig:Cooperation}a. Results show that the greatest level of cooperation is achieved by the \textit{majority-Ut} and \textit{majority-V-Ki} populations, with around 70\% of the players cooperating by the end. In the \textit{majority-V-Ki} population, this high level of cooperation develops early on but then drops to about 60\% around episode 9000, which is then followed by an increase back to 70\% soon after, and a stabilization around episode 13000. An analysis of behavior by each player type (see Appendix,  Figures 7a \& 7b) shows that \textit{V-Ki} players in particular go through an initial period of increasing defection - this coincides with an instability in the whole population where defection is still preferable because it increases the probability of being selected by others. Eventually, many of the players learn to avoid defectors and the \textit{V-Ki} players learn their optimal policy of nearly-full cooperation. 

In the \textit{majority-Ut} population, cooperation emerges more slowly, remaining around 60\% for the first 10000 episodes, but then stabilizes without further drops after episode 13000 (at a similar point in time as the \textit{majority-V-Ki} population). Specifically, \textit{Ut} players take longer than \textit{V-Ki} to learn to cooperate (see Appendix, Figures 7a \& 7b), but once cooperation levels increase to around 70\%, it remains stable.

\begin{figure*}[t]
        \includegraphics[height=0.28\linewidth]{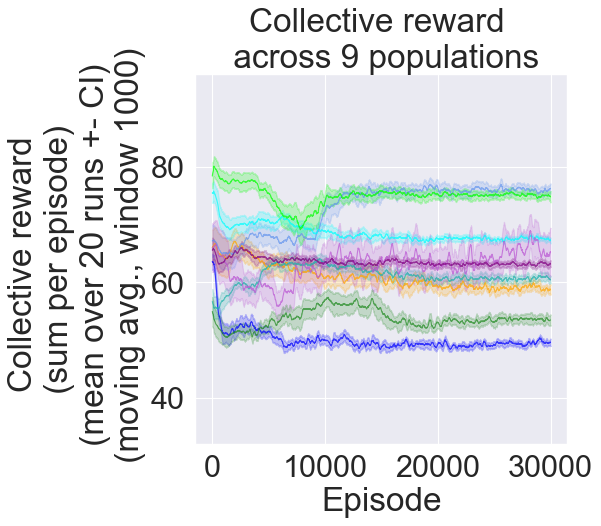}        
        \includegraphics[height=0.28\linewidth]{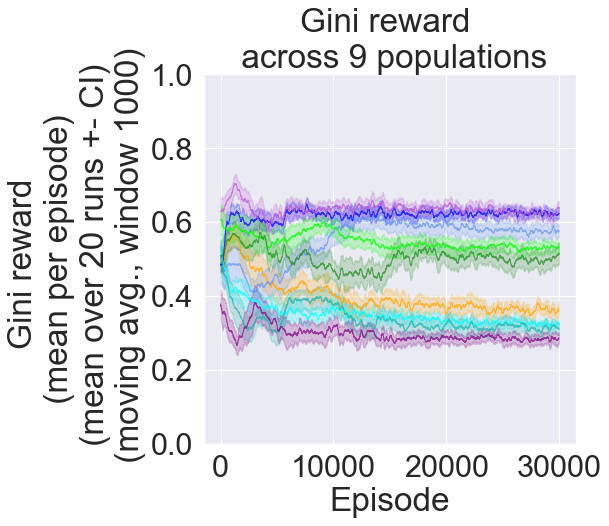}        
        \includegraphics[height=0.28\linewidth]{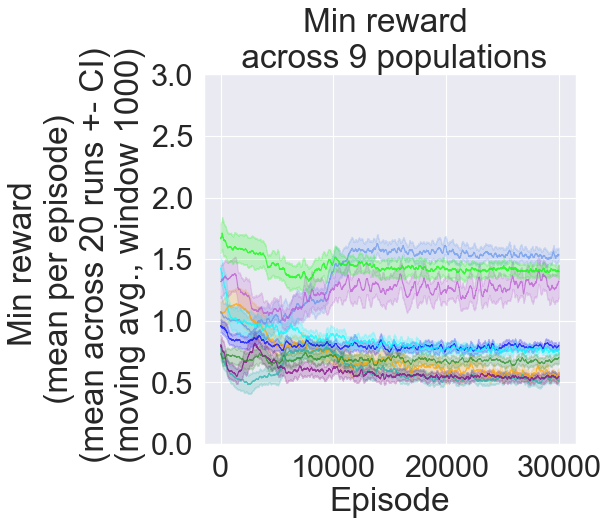}        
        \begin{center} 
        \includegraphics[width=1\linewidth]{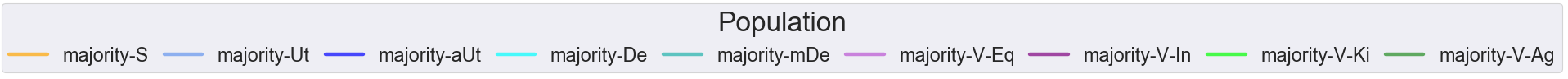}
        \end{center}
    \caption{Population-level social outcomes over time: Collective Reward, Gini Reward and Min Reward. We  plot the moving average of the mean across 20 runs.}
    \label{fig:SocialOutcomes_allplayers}
\end{figure*}

The other majority-prosocial populations \textit{majority-De} and \textit{majority-V-Eq} do not promote as much cooperation. In the \textit{majority-De} population, cooperation drops quickly in the first 1000 episodes and then stabilizes at 60\%  - lower than for \textit{majority-Ut} and \textit{majority-V-Ki} populations. An analysis of action pairs in each population (see Appendix, Figure 6) shows that \textit{De} players in this population learn an near-100\% cooperative strategy very early on. The low level of cooperation in the \textit{majority-De} population can therefore be explained by the behavioral dynamics involving other player types: this population in particular is characterized by high levels of exploitation (see Appendix, Figure 6), whereby the \textit{De} players select aggressive opponents such as \textit{aUt} or \textit{V-Ag}, and those opponents consistently defect against the \textit{De} players, thus bringing the overall level of cooperation in the population down. We discuss the corresponding selection dynamics later in this section.

Finally, both the \textit{majority-V-Eq} and the \textit{majority-V-In} populations achieve similar levels of cooperation, which are lower than for any of the other majority-prosocial population. This can be explained by the fact that both the \textit{V-Eq} and \textit{V-In} players never learn fully cooperative policies (see Appendix, Figure 7b) - due to the structure of the IPD game, where both (C,C) and (D,D) actions lead to equal outcomes. The difference between these populations is that \textit{majority-V-Eq} displays much more instability in terms of cooperation per episode (see Figure \ref{fig:Cooperation}), and more mutual cooperation or defection than the \textit{majority-V-In} population (see Appendix, Figure 6). 

The least amount of cooperation is observed in the \textit{majority-aUt} population. An analysis of pairs of actions over time (see Appendix, Figure 6) shows that this is due to the prevalence of mutual defection between any pair of players in particular, and follows strategically from the fact that mutual defection allows for the lowest collective payoff in the IPD, satisfying the goals of the \textit{aUt} agents.

\subsubsection{Behavior of selfish learners.}

In addition, we analyze the cooperative behavior of \textit{S} learners in each population. These players are the most aligned to traditional multi-agent learning literature, since they do not learn according to any intrinsic reward, but rather by simply maximizing the game reward according to the IPD payoff matrix. This analysis investigates which population compositions steer \textit{S} learners into more cooperative behaviors. An analysis of cooperative behavior exhibited by each non-\textit{S} player type across populations is available in Appendix, Figures 7a \& 7b.

As shown in Figure \ref{fig:Cooperation}b, \textit{S} players generally learn a mostly-defective policy in most populations, though cooperation is slightly above the 5\% chance level even at the end of the simulation, likely due to pressure to cooperate to get selected by some other players. \textit{S} players display most cooperation in the \textit{majority-V-Eq} and the \textit{majority-Ut} populations, though in the latter cooperation still decreases to 10\% towards the second half of the training period. Thus, the \textit{majority-V-Eq} population has a unique influence on the learning of the \textit{S} agent, and is able to steer this agent towards a greater level of cooperation for longer. 
The \textit{V-Eq} opponent is about 40\% likely to cooperate in this population (see Appendix, Figure 7b), which must drive the \textit{S} player to also learn the near-40\% cooperative policy observed here. 

\subsection{Social Outcomes}

In addition to cooperation levels, we analyze a set of social outcomes for each population. We adopt standard social outcome metrics used in the existing literature \citep{hughes2018inequity,tennant2023modeling} - in particular, we measure collective reward ($R_{collective}$), equality between payoffs ($R_{gini}$) and the lowest payoff obtained ($R_{min}$) for a pair of agents $\{M,O\}$ that play on any iteration $t$. 
To aggregate values from every iteration $t$ within an episode (where the length of the episode is equal to the population size $N=16$), we sum $R_{collective}$, and average $R_{gini}$ and $R_{min}$ per episode as follows:
%
\begin{align} 
R_{collective}&= \sum\nolimits_{t=1}^N {(R_{M_{extr}}^t+R_{O_{extr}}^t)} \\
R_{gini}&=\dfrac{ \sum_{t=1}^N {(1-\frac{|R_{M_{extr}}^t-R_{O_{extr}}^t|}{R_{M_{extr}}^t+R_{O_{extr}}^t})} } {N} \\
R_{min}&=\dfrac{\sum_{t=1}^N {\min(R_{M_{extr}}^t,R_{O_{extr}}^t)} } {N}. 
\end{align} 



\begin{figure*}[t]
    \begin{center}
        \includegraphics[width=1.90\columnwidth]{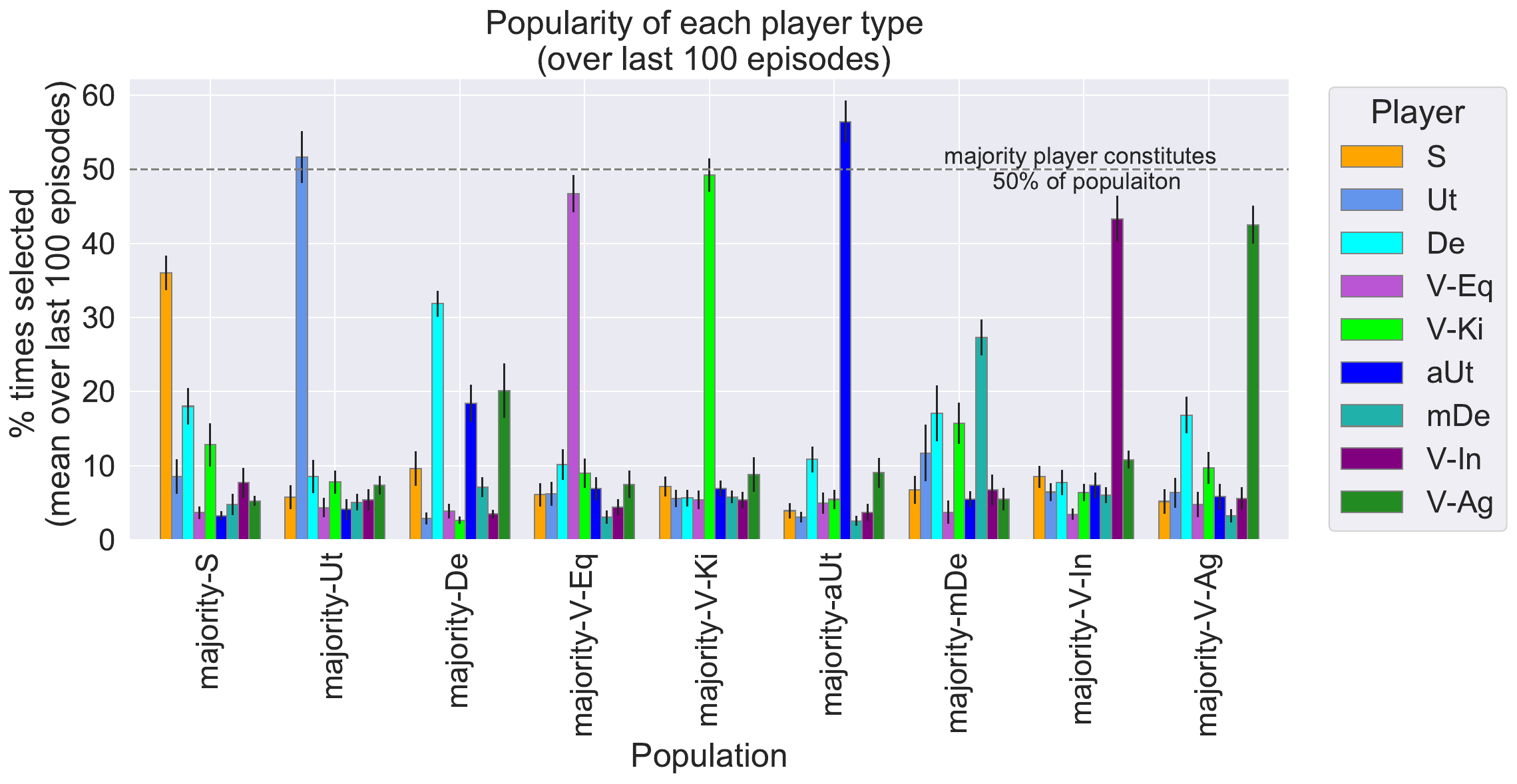}
    \end{center}
    \caption{Popularity of player types in each population on the final 100 episodes. Values represent the average across 20 runs and the associated confidence intervals. We sum the values for cases where more than one player of the same type is present (e.g. 8x\textit{S} players in the \textit{majority-S} population). For ease of interpretation, we add a 50\% reference line - this allows us to compare whether the majority player is selected more (or less) often than expected simply due to their prevalence in each population.}
    \label{fig:Selections_heatmap_a}
\end{figure*}

\begin{figure*}[h]
    \centering
           (a) \includegraphics[height=0.46\linewidth]{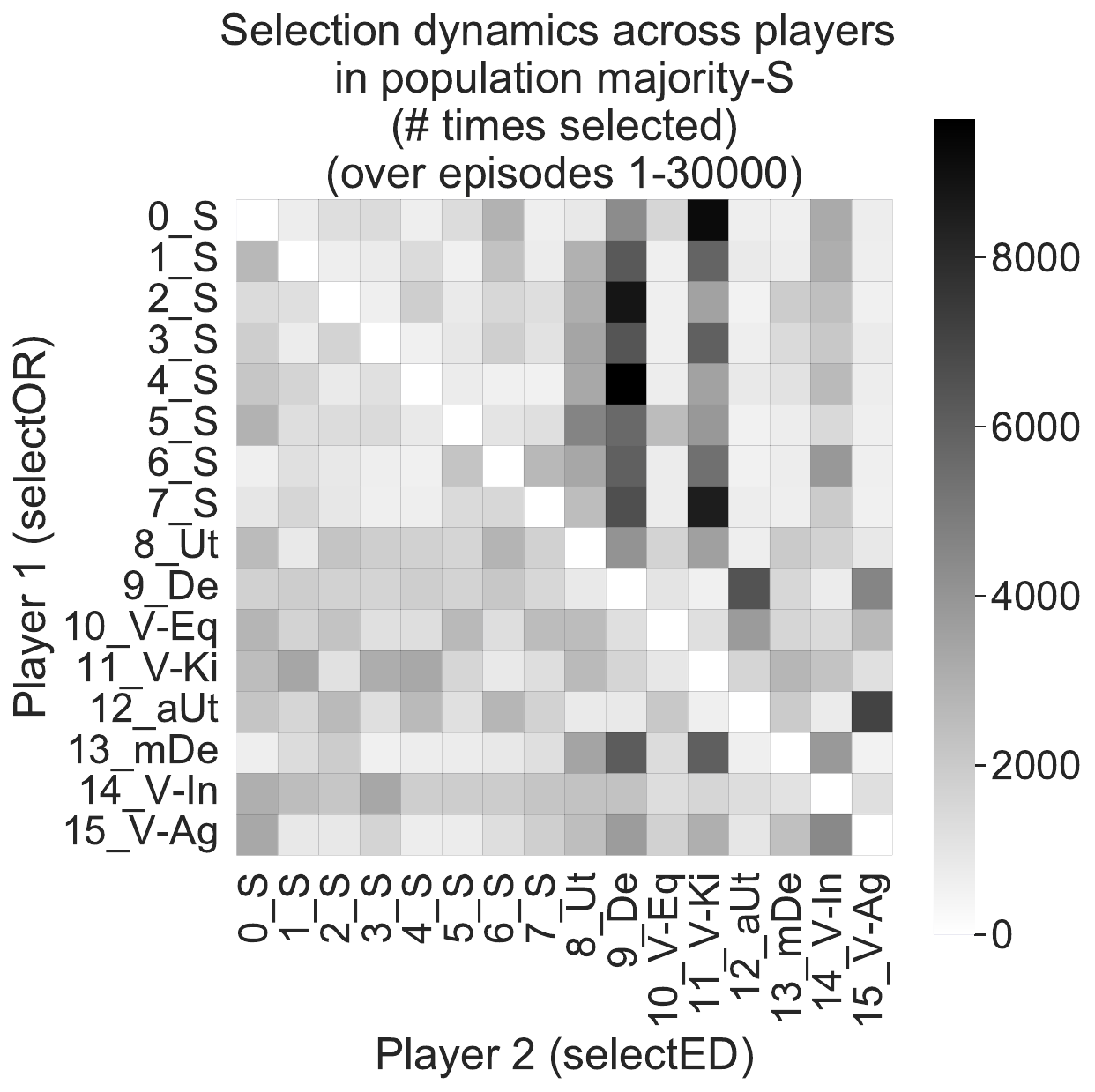}
           (b) \includegraphics[height=0.46\linewidth]{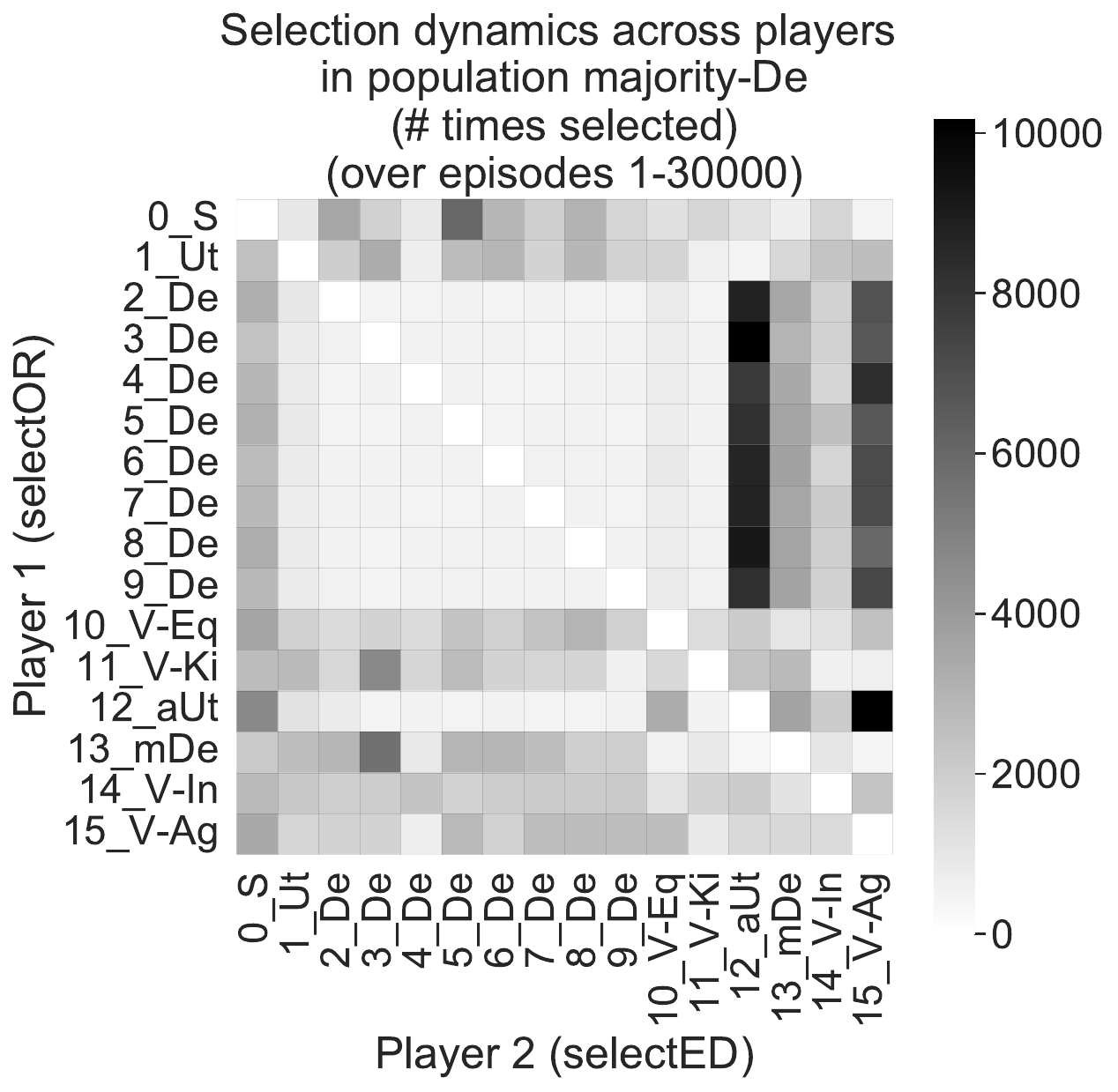}
    \caption{
    Selections made by individual players, with number of selections summed over all 30000 episodes (average across 20 runs), for two example populations: \textit{majority-S} \textbf{(a)} \& \textit{majority-De} \textbf{(b)}; for all populations, see Appendix, Figures 8a, 8b, 9a, 9b.} 
    \label{fig:Selections_heatmap_bc}
\end{figure*}

In Figure \ref{fig:SocialOutcomes_allplayers} we present the outcomes over time in each population. Collective reward follows a similar pattern to population-level cooperation (Figure \ref{fig:Cooperation}a), since these are interconnected in the IPD (in particular, cooperation by one or both of the players on the IPD leads to greater collective reward than mutual defection, as shown by the payoffs in Table \ref{tab:game}). 

Equality between the players (i.e., Gini reward) is highest for the equality-focused \textit{majority-V-Eq} population - though the absolute mean value per episode does not go above $0.7$, meaning that unequal outcomes remain prevalent even in this population. Interestingly, an analysis of action pairs (see Appendix, Figure 6) shows that this level of equality is achieved in this population equally through mutual cooperation and mutual defection. Equality is also generally high for the pro-social populations \textit{majority-Ut} and \textit{majority-V-Ki} due to the prevalence of mutual cooperation, and for their anti-social counterparts \textit{majority-aUt} and \textit{majority-V-Ag} due to the prevalence of mutual defection. Equality is lower when the majority of players are of the \textit{S, De, mDe} or \textit{V-Ie} types, since in these scenarios one player tends to exploit 
the other. 

Minimum reward is highest once again in the \textit{majority-Ut} and the \textit{majority-V-Ki} populations (with values around $1.5$, due to the prevalence of mutual cooperation - see Appendix, Figure 6), and also in the \textit{majority-V-Eq} case (due to a lack of exploitation). All other populations result in lower minimum reward of values between $0.5$ and $1.0$, which are explained by either exploitation of mutual defection in the IPD.

\subsection{Selection Dynamics}\label{subsec:Selection}

Given the selection mechanism implemented, it is interesting to analyze which players are the most popular overall in each population (Figure \ref{fig:Selections_heatmap_a}), and which selector types prefer which types of partners (Figure \ref{fig:Selections_heatmap_bc}). In a population of purely selfish agents \citep{Anastassacos2020partner}, the selection mechanism adds an incentive to cooperate, since selfish players benefit from playing against cooperators and tend to select them. In our heterogeneous populations, however, certain players may not necessarily prefer cooperative opponents, so selection dynamics are likely to be more complex.

Figure \ref{fig:Selections_heatmap_a} shows the popularity of each player type in each population over the final 100 episodes. Generally, in probabilistic terms, especially at the beginning, the majority player type tends to be the most popular in their respective population. However, we observe that in certain populations, for example \textit{majority-S} and \textit{majority-De}, the majority player type is selected less than 50\% of the time - i.e., much less frequently than expected due to their prevalence. In these cases, it is interesting to analyze which non-majority players emerge as popular alternatives. For the full analysis of selection patterns across individual players in all populations, see Appendix, Figures 8a, 8b, 9a, 9b.

The behavior of the \textit{majority-De} population is the most striking. Here, \textit{aUt} and \textit{V-Ag} players are relatively popular. A further analysis of the specific selection patterns among individual players in this population (Figure \ref{fig:Selections_heatmap_bc}b) reveals that the \textit{De} players in particular tend to mostly prefer anti-social opponents \textit{aUt} and \textit{V-Ag}. This apparently self-sabotaging behavior can be explained by a closer analysis of the \textit{De} player's $R_{intr}$ - they are penalized for defecting against a cooperator, so the safest behavior in a heterogeneous population is to always select a defector, so that they have the lowest chance of violating their moral norm. Figure \ref{fig:Selections_heatmap_bc}b also shows that another player - \textit{aUt} - prefers to avoid the \textit{De} majority players and instead tends to select the \textit{V-Ag} opponent. This highlights that minimizing collective reward (which is the preference embedded in the \textit{aUt} agent's intrinsic reward) is best achieved by avoiding the very cooperative \textit{De} agent. Thus, we find an interesting dynamic in the heterogeneous \textit{majority-De} population, in which the majority of the players follow a strong cooperative norm, demonstrating ways in which these \textit{De} players can be very cooperative (see Figure \ref{fig:Cooperation}a and Appendix, Figure 6) but not very popular. 

Returning to the analysis of player popularity on the final 100 episodes (Figure \ref{fig:Selections_heatmap_a}), the \textit{majority-mDe} population also has a few non-majority players emerging as popular. In this case, it is the \textit{De} and \textit{V-Ki} players that get selected often by others. Thus, in both \textit{majority-De} and \textit{majority-mDe} populations, the norm-following majority player is not even popular among their own kind.

\begin{figure*}[h]
    \begin{center}
        \includegraphics[height=0.40\linewidth]{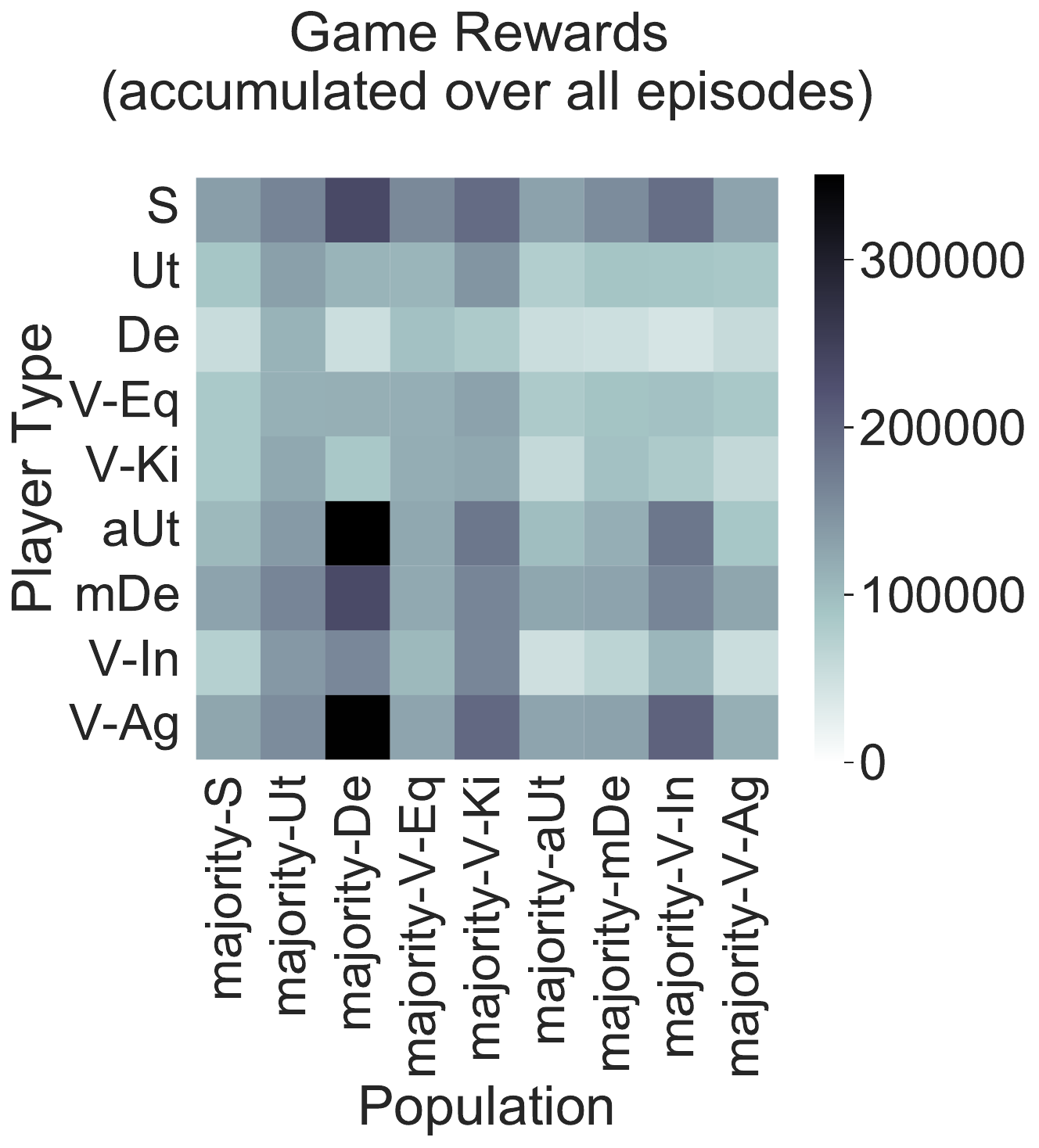}
        \includegraphics[height=0.40\linewidth]{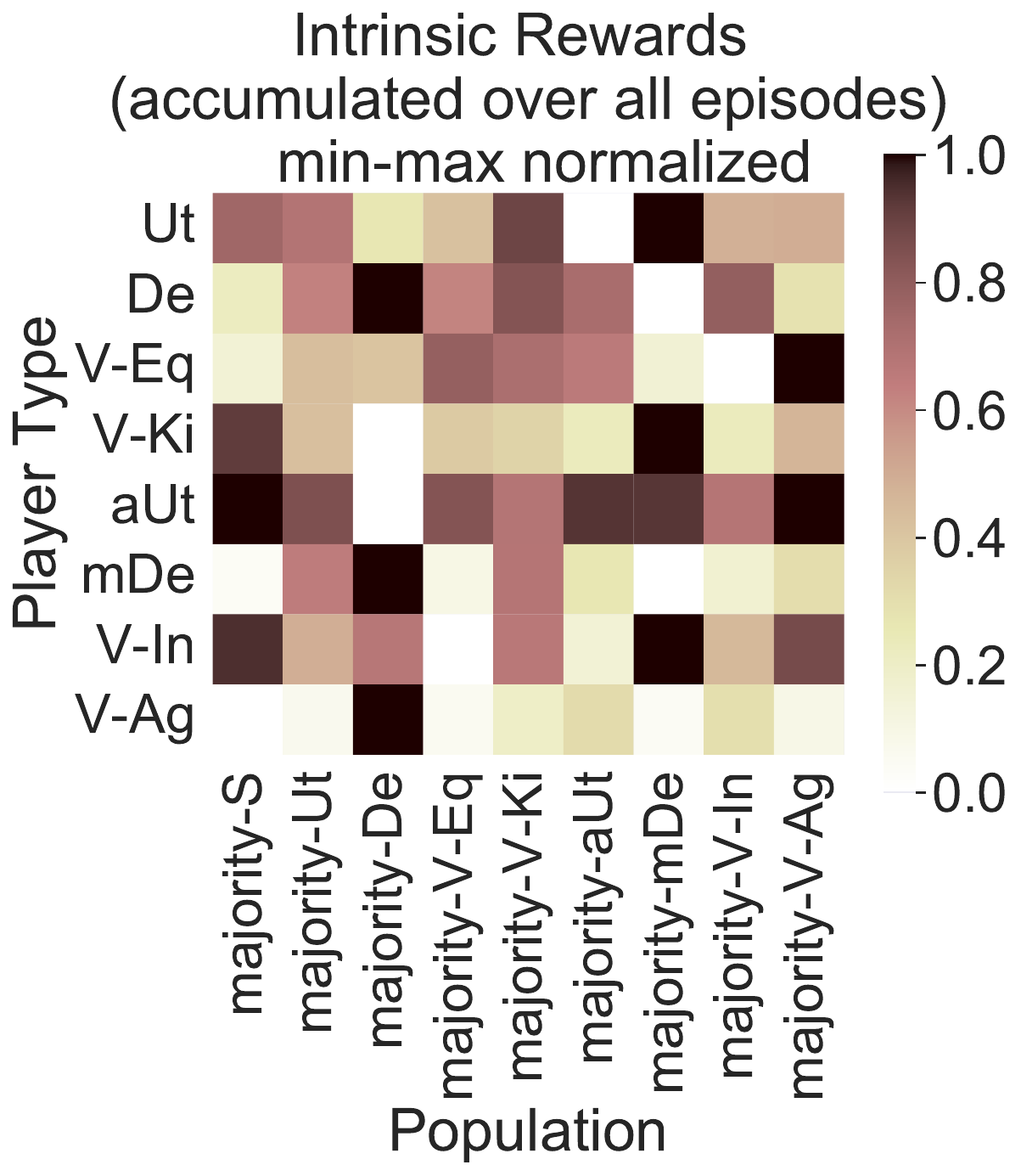}
    \end{center}
    \caption{Game reward and intrinsic reward accumulated by each player type in each population over the entire 30000 episodes (averaged across 20 runs). We average per number of players of a certain type (e.g. 8x\textit{S} players in the \textit{majority-S} population). For comparability, we normalize intrinsic rewards using the minimum and maximum observed value for each player type.}
    \label{fig:GameReward}
\end{figure*}

Finally, the \textit{majority-S} population is also interesting to analyze (Figure \ref{fig:Selections_heatmap_bc}a), because this population differs from others in that most agents here are pure payoff-maximizers in terms of the traditional IPD game. One might expect that \textit{S} players would often select \textit{De} or \textit{V-Ki} opponents, since these are so easily exploitable, but in fact \textit{Ut} and \textit{V-In} also emerge as relatively popular, likely because these players will have learned to cooperate against a defector, which benefits the \textit{S} player in terms of reward. An analysis of action pairs (see Appendix, Figure 6) confirms that unilateral defection is indeed common in this population. Many of the other players in this population tend to select majority type \textit{S} opponents roughly as often as other player types, meaning that \textit{Selfish} players do not emerge as unpopular, despite their often defective behavior. The only exceptions to this are \textit{mDe} and \textit{V-Ag}. The \textit{mDe} player in particular learns to avoid the \textit{S} opponent since \textit{mDe} would rather face a cooperator (such as \textit{De} or \textit{V-Ki}), and obtain positive $R_{intr}$, than select the \textit{S} players which are likely to defect. 


\subsection{Cumulative Reward}\label{subsec:Individual}

Finally, we analyze the cumulative reward obtained for each individual player type in each population. Figure \ref{fig:GameReward} presents the reward accumulated by each player type in each population over all 30000 episodes. The results in terms of game reward clearly show that - in general - pro-social players are disadvantaged on the IPD game itself, since they get exploited and/or do not benefit from exploiting a cooperative opponent. Furthermore, the lowest-performing player type is \textit{De}. However, the \textit{De} player obtains high levels of intrinsic reward in many scenarios, especially when their own kind constitute the majority of the population. Other pro-social players obtain the highest $R_{intr}$ in populations in which some anti-social player type constitutes the majority. 

Anti-social or \textit{S} players obtain higher game reward over time than their pro-social counterparts - likely by exploiting others. The \textit{majority-De} population allows \textit{aUt} and \textit{V-Ag} players to obtain especially high levels of game reward; given the selection dynamics analyzed above, we understand that this is driven by the fact that \textit{De} players actively prefer to select these opponents and then get exploited by them. The lowest-performing players in terms of intrinsic reward in general are the norm-based anti-social players \textit{mDe} and \textit{V-Ag} (especially in majority-anti-social populations) and the norm-based prosocial players \textit{majority-V-Ki} (especially in majority-pro-social populations). Players of the \textit{aUt} type obtain high intrinsic reward in most populations except \textit{majority-De} (because the presence of such cooperative players makes it hard to minimize collective reward), whereas the \textit{mDe} and \textit{V-Ag} obtain the highest intrinsic reward by exploiting the very cooperative \textit{De} players in the \textit{majority-De} population.

\section{Discussion}

In this paper we have investigated how the presence of different moral preferences in populations affects individual agents' learning behaviors and emergent social outcomes. The results provide insights into behaviors emerging from multi-agent interactions between morally heterogeneous agents. Some of our findings point to the dangers of designing artificial agents in certain ways - for example, showing how this may lead to the development of self-sabotaging behaviors. Additionally, while our study can be considered as a stark simplification of reality, the insights deriving from it can provide a starting point towards investigating the emergence of similar behaviors in human societies.
In summary, many of the agents' actions are consistent with their reward definitions - in general, pro-social agents prefer to cooperate (except for the \textit{V-Eq} agent), and anti-social agents display lower levels cooperation and obtain lower collective reward (except for the \textit{V-In} agent). 

In terms of temporal dynamics, our results demonstrate interesting asymmetries in the emergence of cooperation by agents with different moral preferences. In particular, we observe that consequentialist agents \textit{Ut} take longer to learn to cooperate than the norm-based agents \textit{De}, while the norm-based agents \textit{V-Ki} go through an unstable and defective period before learning a stable cooperative policy. Furthermore, for certain consequentialist agents (specifically, \textit{aUt} vs \textit{Ut}), the convergence to an equilibrium in the Prisoner's Dilemma environment is faster for anti-social players than their pro-social counterparts - likely due to the payoffs attributed to defection in this game.

We have also observed some surprising interactions between different types of morality due to the selection mechanism present.
First, the introduction of a large number of equality-focused agents \textit{(V-Eq)} has a positive effect on the cooperative behavior of a \textit{Selfish} agent \textit{(S)}, providing insight into how diverse opponent types can steer self-interested agents towards more pro-social behavior. Second, one pro-social player type (specifically, the norm-based player \textit{De}) learns to select anti-social opponents in order to avoid violating their own moral norm. Many anti-social players simultaneously learn to select the \textit{De} player and exploit them in the game, since this exploitative behavior is not penalized by anyone in the population and does not deter the \textit{De} player from further interaction. Thus, our results show that the introduction of \textit{De} type agents risks promoting anti-social actors and inequality in a population. This illustrates the possibility that the presence of narrowly-defined norms might lead to self-sabotaging behavior and cause negative outcomes for the population as a whole.

\section{Conclusion}

This study is the first to analyze the dynamics of learning in populations of agents with moral preferences varying from those with consequentialist foundations to those with norm-based ones. Our results demonstrate the potential of using intrinsic rewards for modeling various moral preferences in RL agents. 
More generally, we have provided a generic methodology for studying the learning dynamics of heterogeneous populations, and introduced measures for assessing individual and societal outcomes. Finally, this work might also contribute to raising awareness about emergent behaviors and the possibility of unintuitive outcomes in such multi-agent learning scenarios.

 Our research agenda includes the study of more complex moral frameworks and an exploration of multi-objective approaches that combine moral and self-interested motivation. We also plan to investigate environments characterized by learning dynamics under partial observability.

\section{Acknowledgments}

Elizaveta Tennant was supported by the Leverhulme Trust (DS-2017-026; Doctoral Training Programme for the Ecological Study of the Brain).
Mirco Musolesi was supported by the Italian Ministry of University and Research (MUR) through the project PRIN 2022 “Machine-learning based control of complex multi-agent systems for search and rescue operations in natural disasters (MENTOR)” funded by the European Union - NextGenerationEU.

\section{Ethical Considerations}

As part of our study, we have designed malicious learning agents, which are able to take advantage of certain exploitable pro-social agents. Studying the effect of the presence of these agents might increase our understanding of the resilience and stability of systems even in presence of malicious actors. 

\bibliography{paper}


\newpage 

\newpage

\appendix

\begin{center}
\textbf{\Large{Appendix}}
\end{center}

\section{Action Pairs within Each Population}
\label{sec:appendix_actionpairs}

An analysis of action pairs in Figure 6 provides further insight into the pair-wise interactions in each population, including mutual cooperation (C,C); unilateral exploitation (C,D or D,C); and mutual defection (D,D) observed at every episode. We observe that the high levels of cooperation in the \textit{majority-Ut} and \textit{majority-V-Ki} populations are largely driven by mutual cooperation, and that the low cooperation in the \textit{majority-aUt}, \textit{majority-V-Ag} and the \textit{majority-V-Eq} populations can be largely attributed to high mutual defection. In other populations, a much higher level of exploitation (i.e., unilateral defection) is observed, and in certain pro-social populations, such as \textit{majority-De}, this is largely due to the select\textit{ing} player being exploited by their select\textit{ed} opponent.

\begin{figure*}[h]
    \begin{center}
        \includegraphics[width=0.32\linewidth]{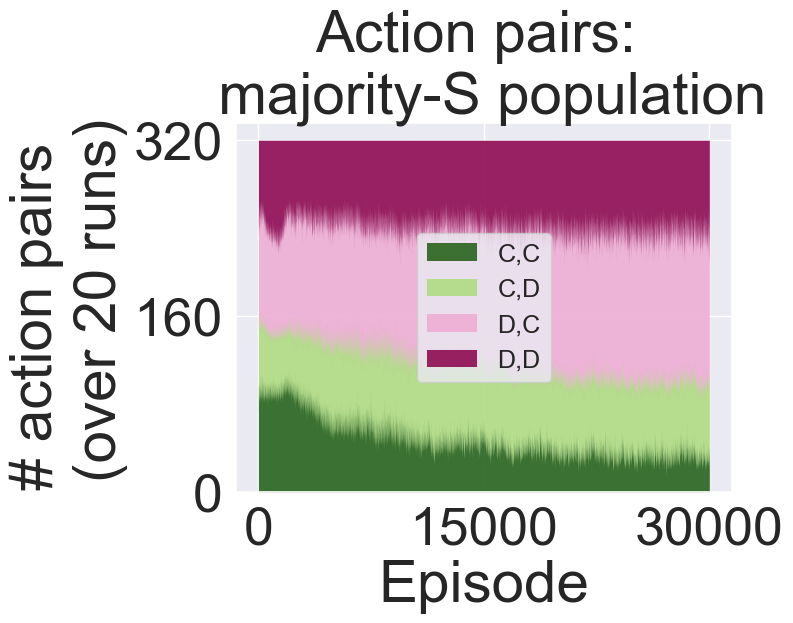}
        \\
        \includegraphics[width=0.32\linewidth]{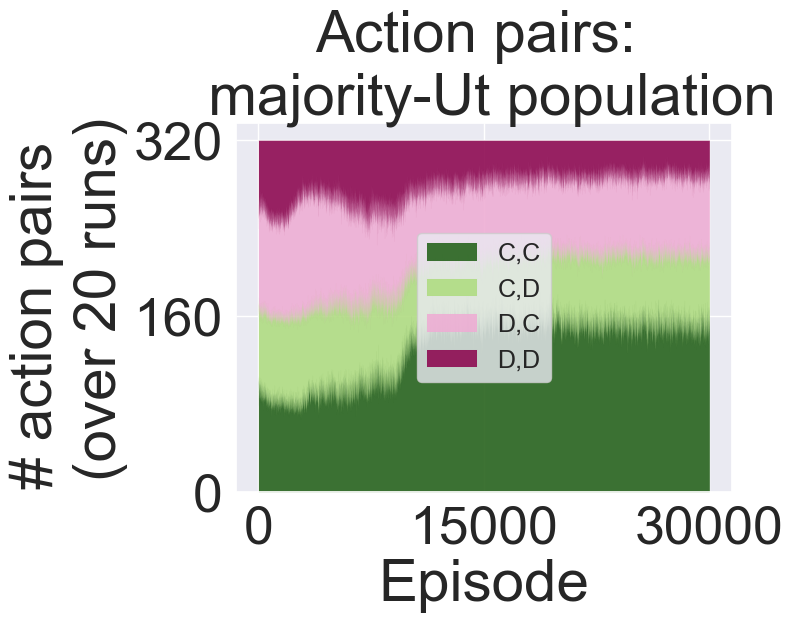}
        \includegraphics[width=0.32\linewidth]{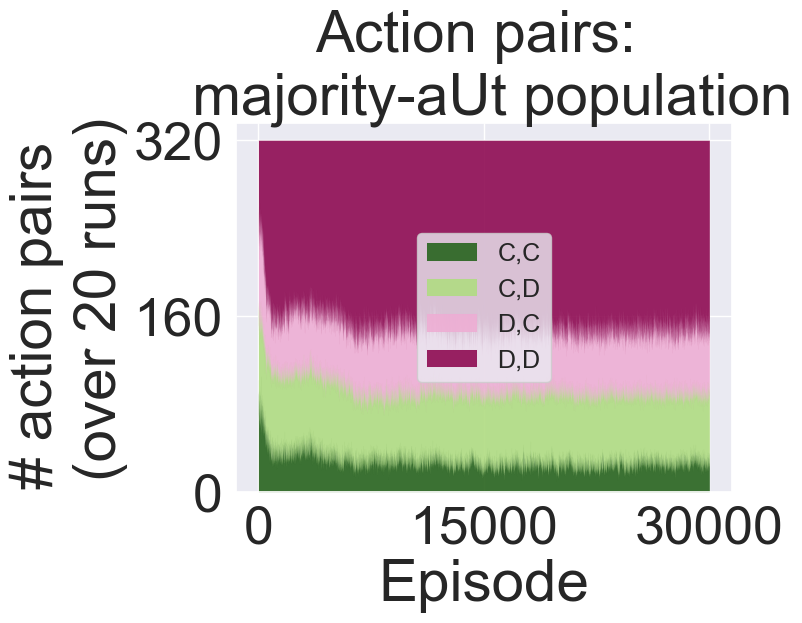}
        \\
        \includegraphics[width=0.32\linewidth]{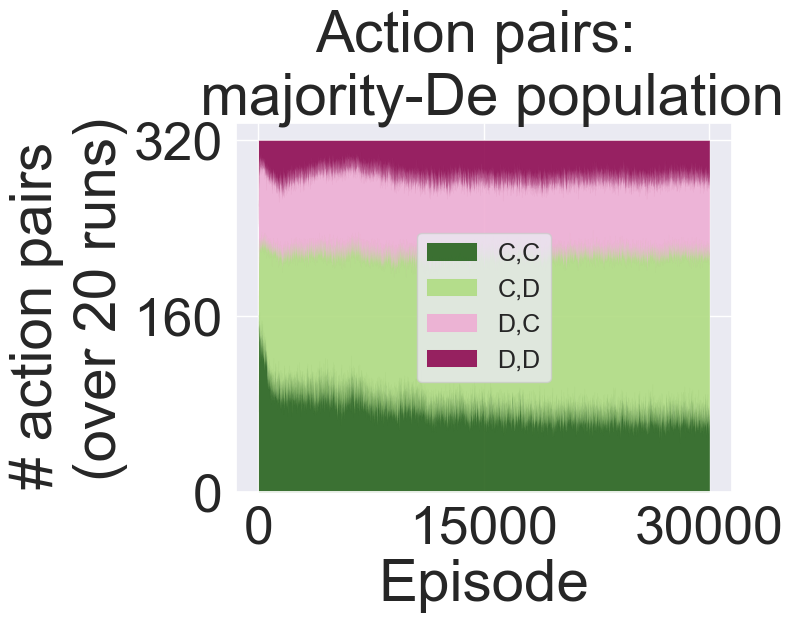}
        \includegraphics[width=0.32\linewidth]{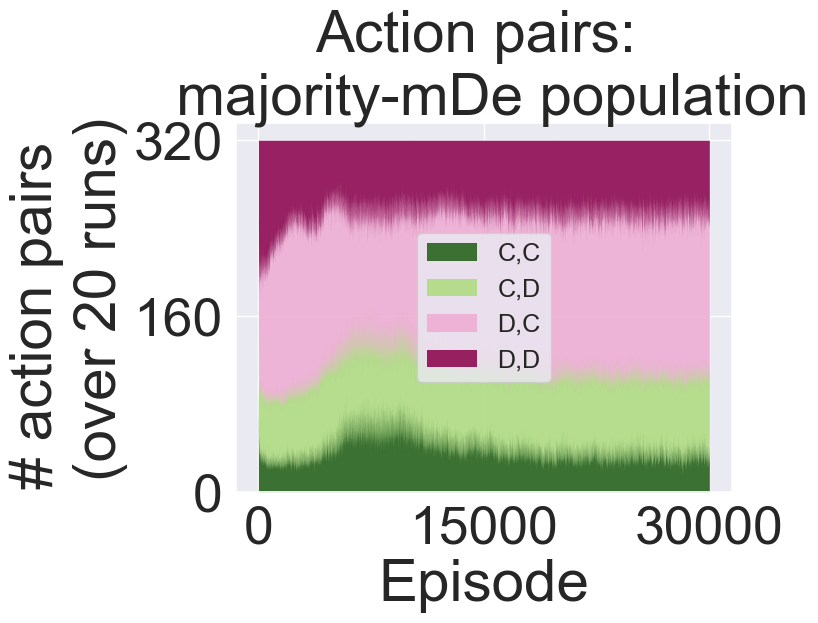}
        \\
        \includegraphics[width=0.32\linewidth]{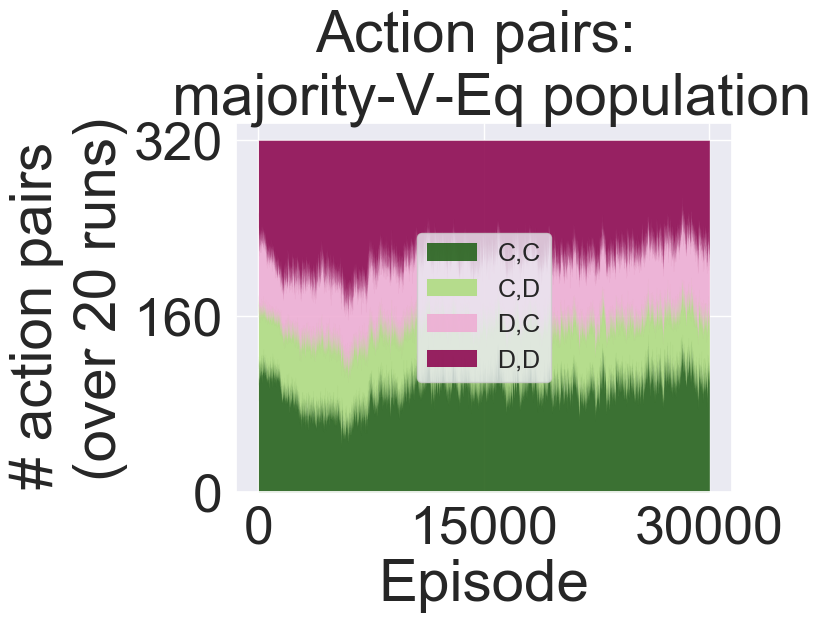}
        \includegraphics[width=0.32\linewidth]{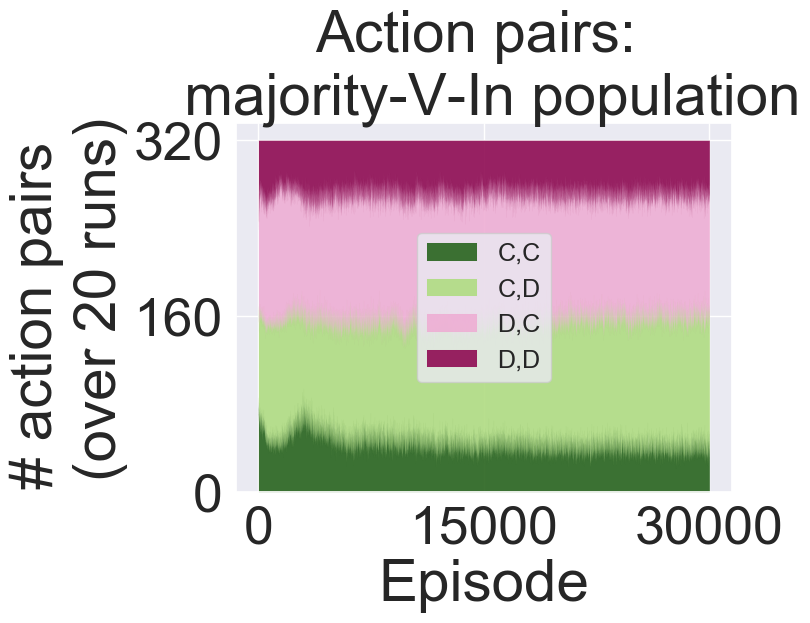}
        \\
        \includegraphics[width=0.32\linewidth]{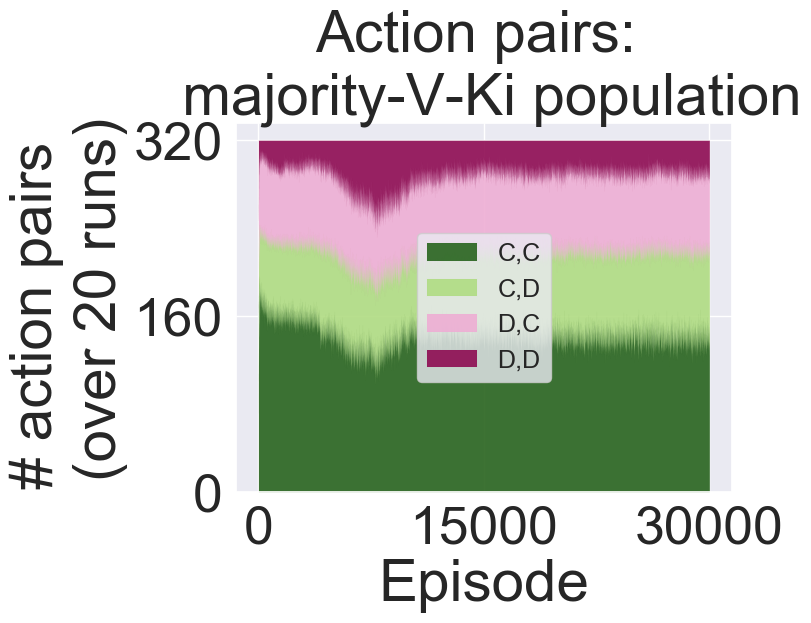}
        \includegraphics[width=0.32\linewidth]{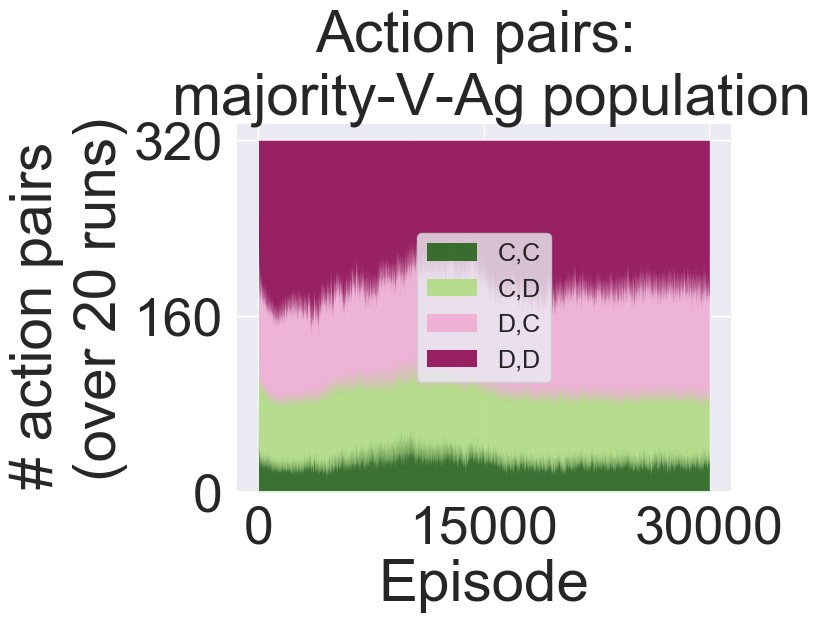}

    \end{center}
    \caption{Action pairs observed on every episode in every population (we sum the 16 interactions within an episode, and then count the number of occurrences of each action pair across 20 runs). The ordering of actions within each pair corresponds to the select\textit{ing} player first, and the select\textit{ed} opponent second. Thus, action pairs are interpreted as follows: (C,C) = mutual cooperation; (C,D) = selecting player gets exploited; (D,C) = selected player gets exploited; (D,D) = mutual defection. }
    \label{fig:ActionPairs}
\end{figure*}

\section{Cooperation by Each Player Type across Populations}
\label{sec:appendix_cooperationbyplayers}

In addition to the analysis of the \textit{S} agents' behavior across populations presented in the main body of the paper, Figures 7a \& 7b present the dynamics of cooperation over time for every other player type in every population.

\begin{figure*}[h]
    \begin{subfigure}{1\textwidth}
        \begin{center}
        \includegraphics[width=0.49\linewidth]{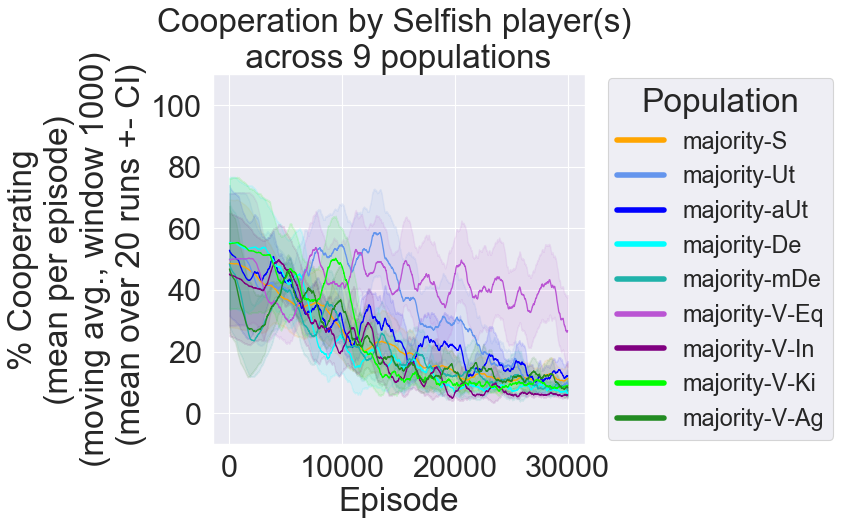}
        \\ 
        \includegraphics[width=0.49\linewidth]{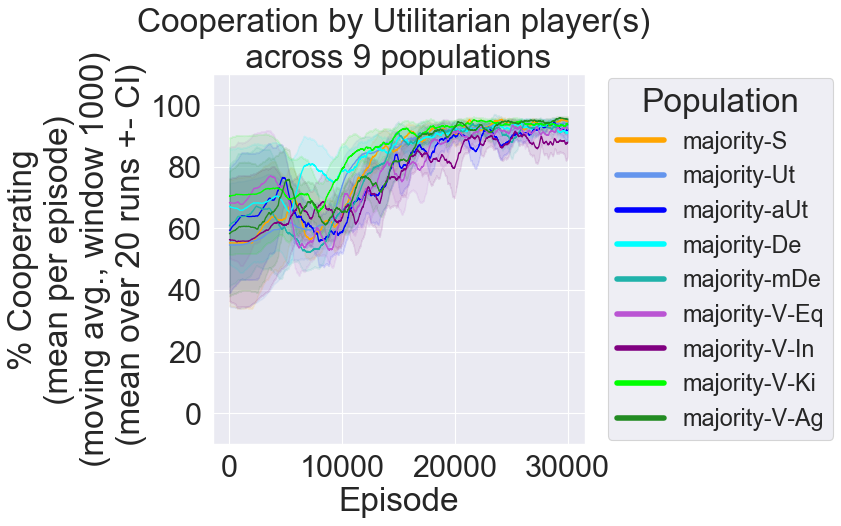}
        \includegraphics[width=0.49\linewidth]{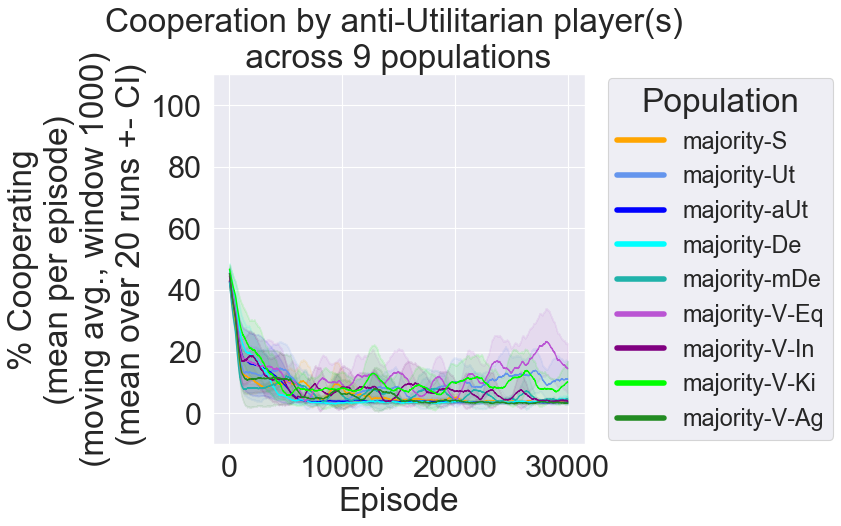} 
        \includegraphics[width=0.49\linewidth]{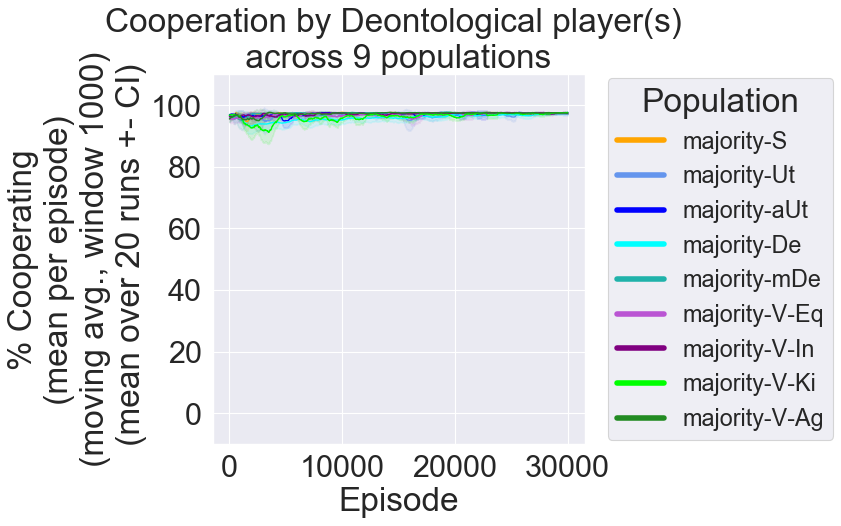}
        \includegraphics[width=0.49\linewidth]{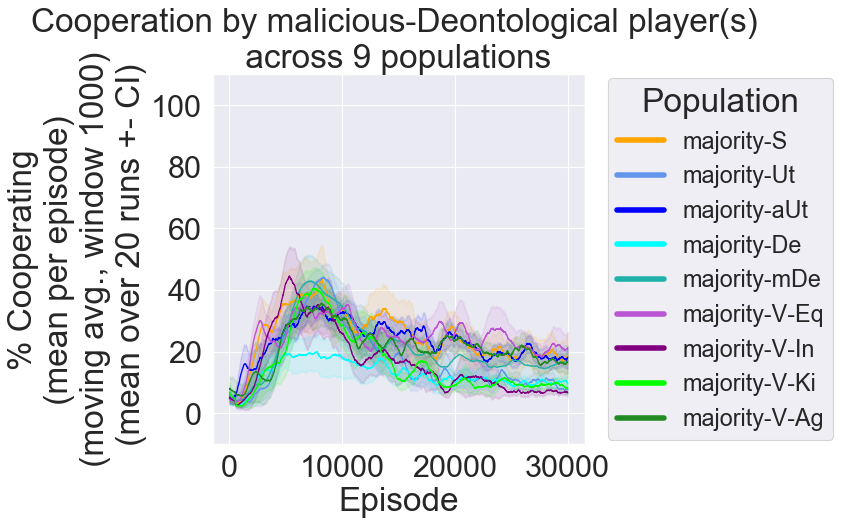}
        \end{center}
    Figure 7a: Cooperation by each player type over time in populations \textit{majority-S, majority-Ut, majority-aUt, majority-De} and \textit{majority-mDe}. In these charts, we plot the moving average of the mean across 20 runs. Where more than one player of a certain type is present in a population, we average the values across all the players of that type (e.g. 8x\textit{S} players in the \textit{majority-S} population).
    \label{fig:sub1}
    \end{subfigure}

\end{figure*}

\begin{figure*}[h]
    \begin{subfigure}{1\textwidth}
        \begin{center}
        \includegraphics[width=0.49\linewidth]{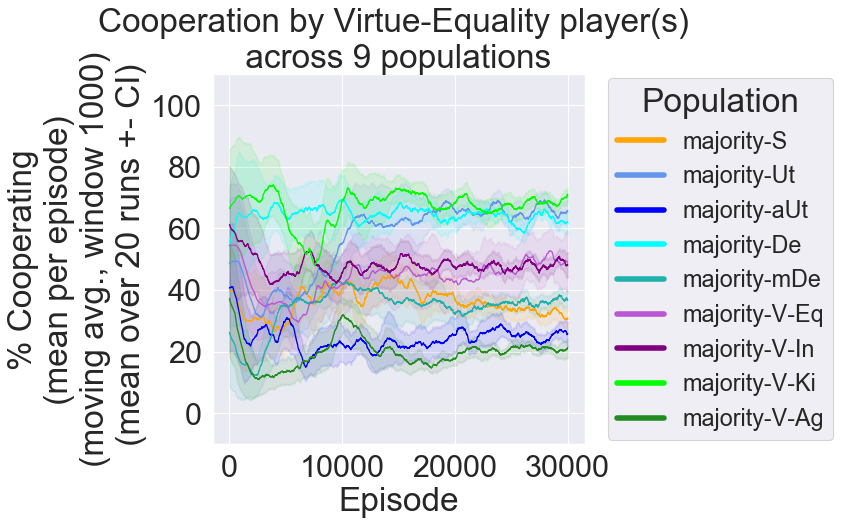}
        \includegraphics[width=0.49\linewidth]{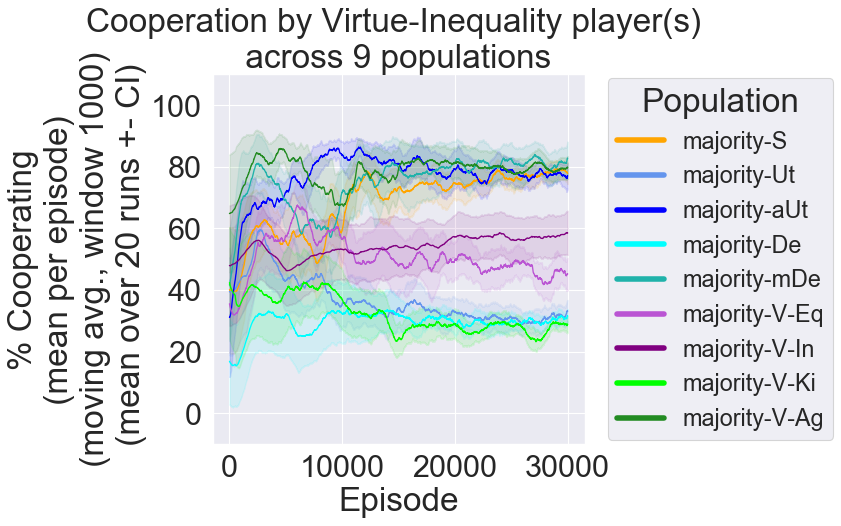}
        \includegraphics[width=0.49\linewidth]{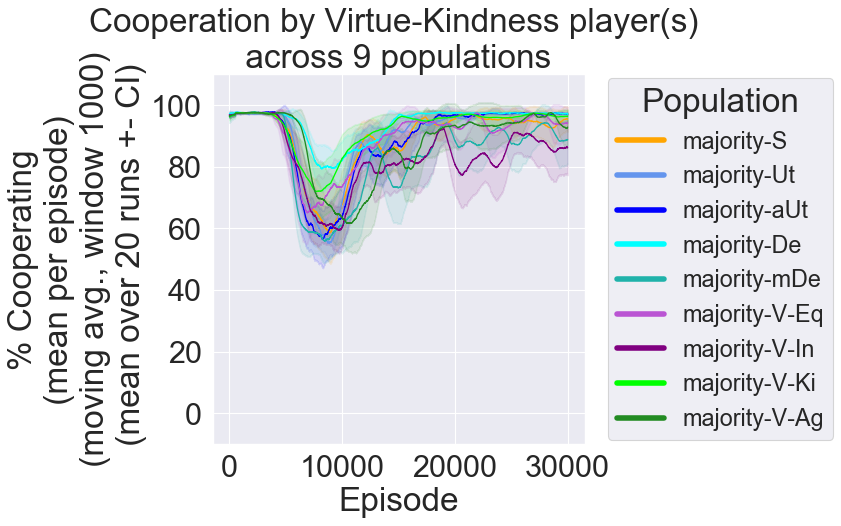}
        \includegraphics[width=0.49\linewidth]{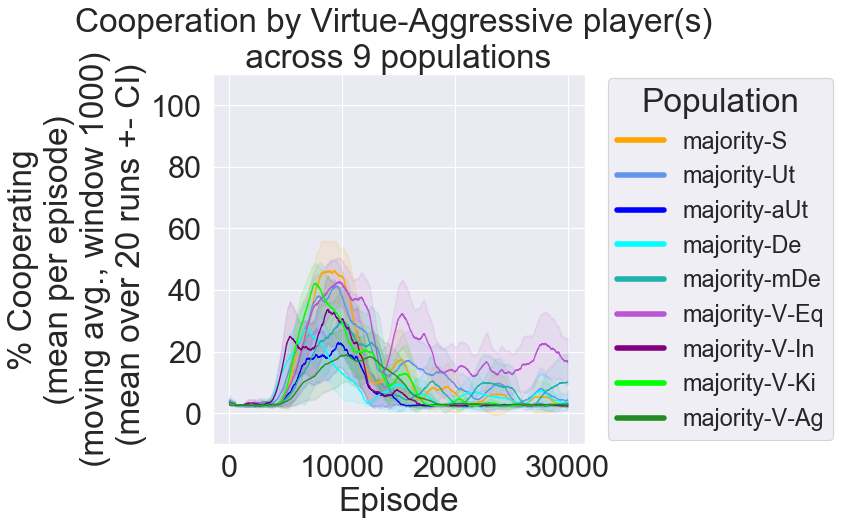}
        \end{center}
    Figure 7b: Cooperation by each player type over time in populations \textit{majority-V-Eq}, \textit{majority-V-In}, \textit{majority-V-Ki} and \textit{majority-V-Ag}. In these charts, we plot the moving average of the mean across 20 runs. Where more than one player of a certain type is present in a population, we average the values across all the players of that type (e.g. 8x\textit{S} players in the \textit{majority-S} population).
    \label{fig:sub2}
    \end{subfigure}
    \label{fig:Eahcplayer_cooperation}
\end{figure*}

For all players except \textit{V-Eq} and \textit{V-In}, levels of cooperation are relatively consistent across populations. \textit{V-Ki} and \textit{V-Ag} players go through an initial phase of low or high cooperation, respectively, after which we observe stable behaviors consistent with their intrinsic reward (i.e., cooperation by \textit{V-Ki} and defection by \textit{V-Ag}).
The other norm-based pro-social player \textit{De} converges to full cooperation very quickly, showing that this definition of the norm produces the strongest signal for cooperation in the IPD (extending the findings of \citealt{tennant2023modeling} to the population case as well). This agent's anti-social counterpart \textit{mDe}, however, does not display full defection, likely due to the partner selection mechanism providing an incentive to cooperate in order to get selected more often to play. The \textit{Ut} and \textit{aUt} players converge to either full cooperation or full defection respectively in all populations nearly equally, but the rate of convergence is faster for the anti-social player, likely due to the payoffs associated with defection in the IPD game. The \textit{V-Eq} and \textit{V-In} players cooperate to variable extents in different populations, but in a way that is consistent with the choice of their reward functions (e.g., defect more in a majority-defective population).

\section{Selection Patterns between Player Types}
\label{sec:appendix3}

In addition to the analysis of selections made over the entire simulation in the two populations presented in the main body of the paper (\textit{majority-S} and \textit{majority-De}, see Figure \ref{fig:Selections_heatmap_bc}a \& \ref{fig:Selections_heatmap_bc}b), we show selections made by each individual player in every population in Figures 8a \& 8b (as a heat map) and Figures 9a \& 9b (as networks involving the top 15\% of the pairs in terms of number of interactions observed). The latter show the most common patterns of interaction among players in these artificial societies - for example, in the \textit{majority-De} population, the \textit{De} players select other types of opponents more often than each other. 

\begin{figure*}[h]
    \begin{center}
        \includegraphics[height=0.45\linewidth]{plots/selections_byplayer/matrix_selections_bipartite_byplayertype_populmajority-S_thresholdFalse.pdf}
        \\ 
        \includegraphics[height=0.45\linewidth]{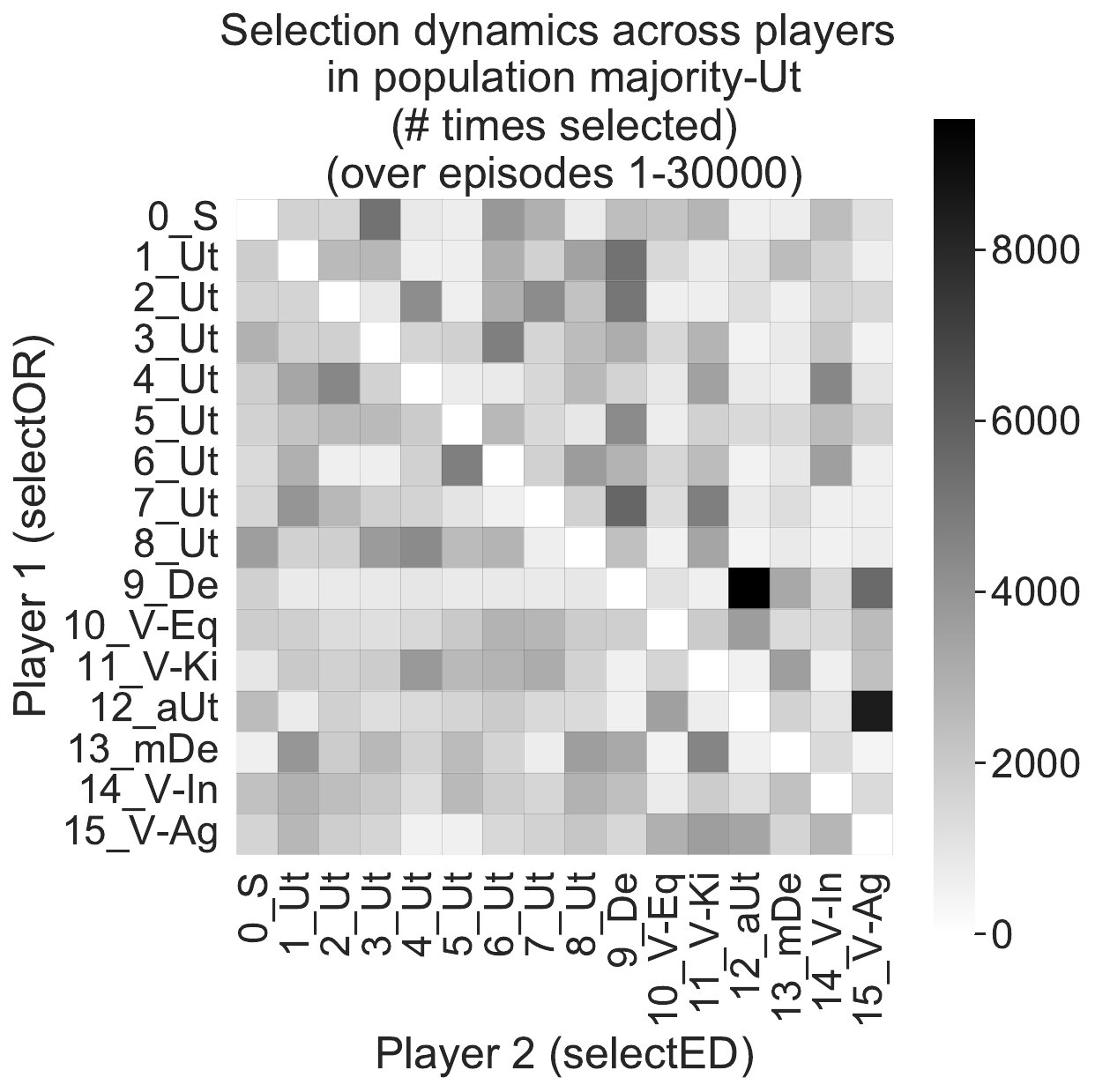}
        \includegraphics[height=0.45\linewidth]{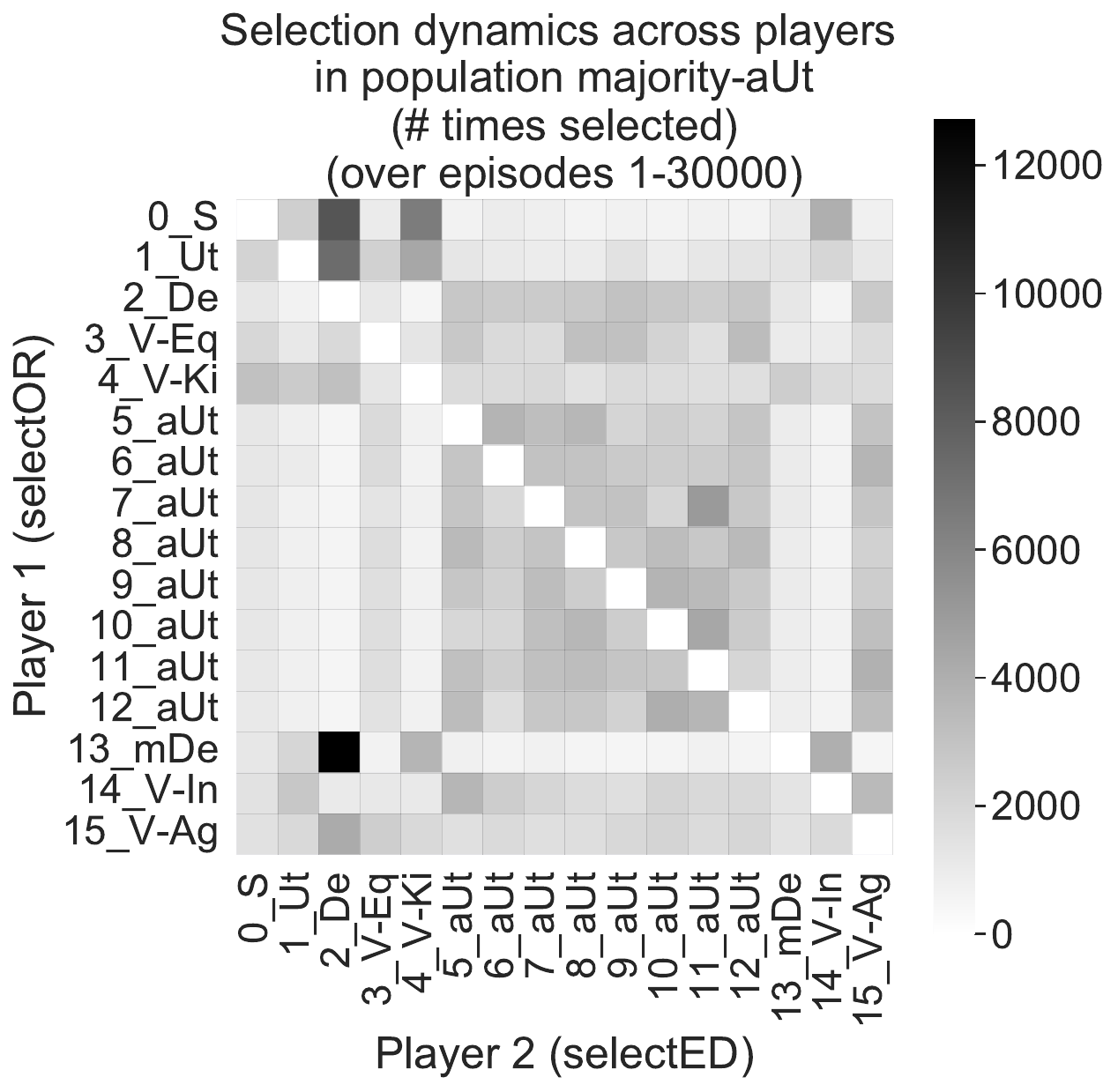}
        \\
        \includegraphics[height=0.45\linewidth]{plots/selections_byplayer/matrix_selections_bipartite_byplayertype_populmajority-De_thresholdFalse.pdf}
        \includegraphics[height=0.45\linewidth]{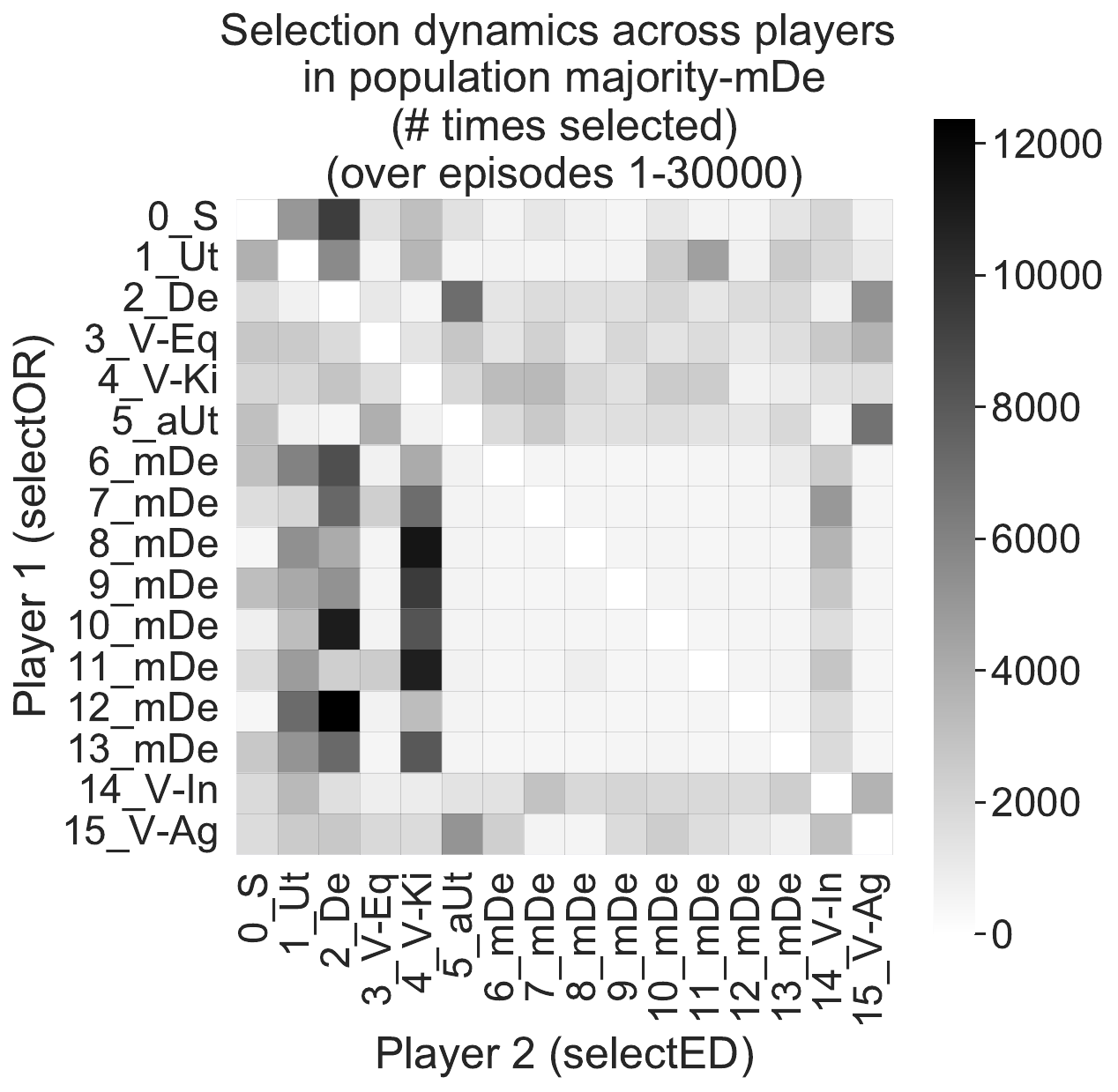}
        
    \end{center}
    Figure 8a: Selections made by individual player types in populations \textit{majority-S, majority-Ut, majority-aUt, majority-De} and \textit{majority-mDe}. The number of selections in every cell is summed over all 30000 episodes (average across 20 runs).
\end{figure*}

\begin{figure*}[h]
    \begin{center}
        
        \includegraphics[height=0.45\linewidth]{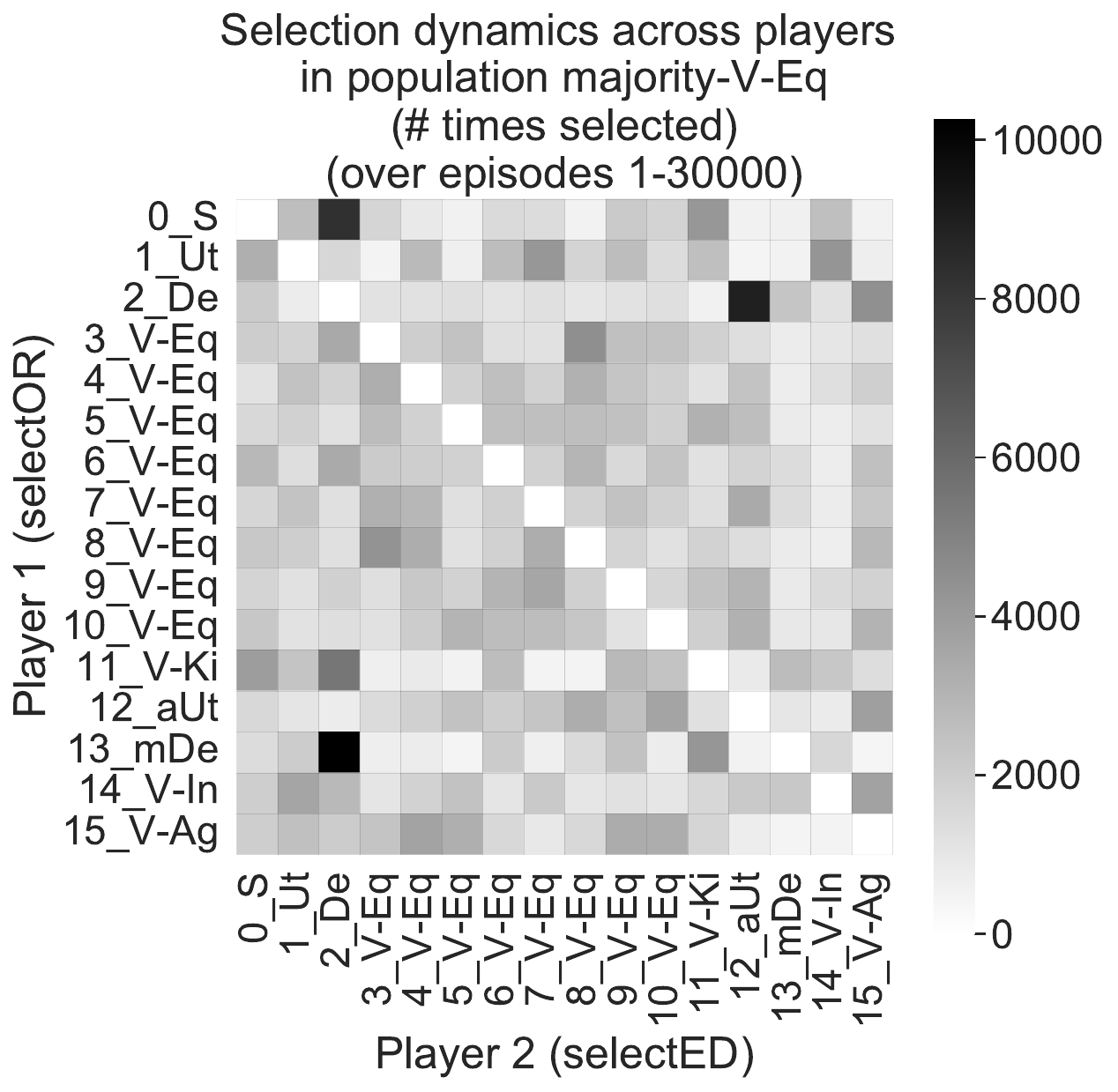}
        \includegraphics[height=0.45\linewidth]{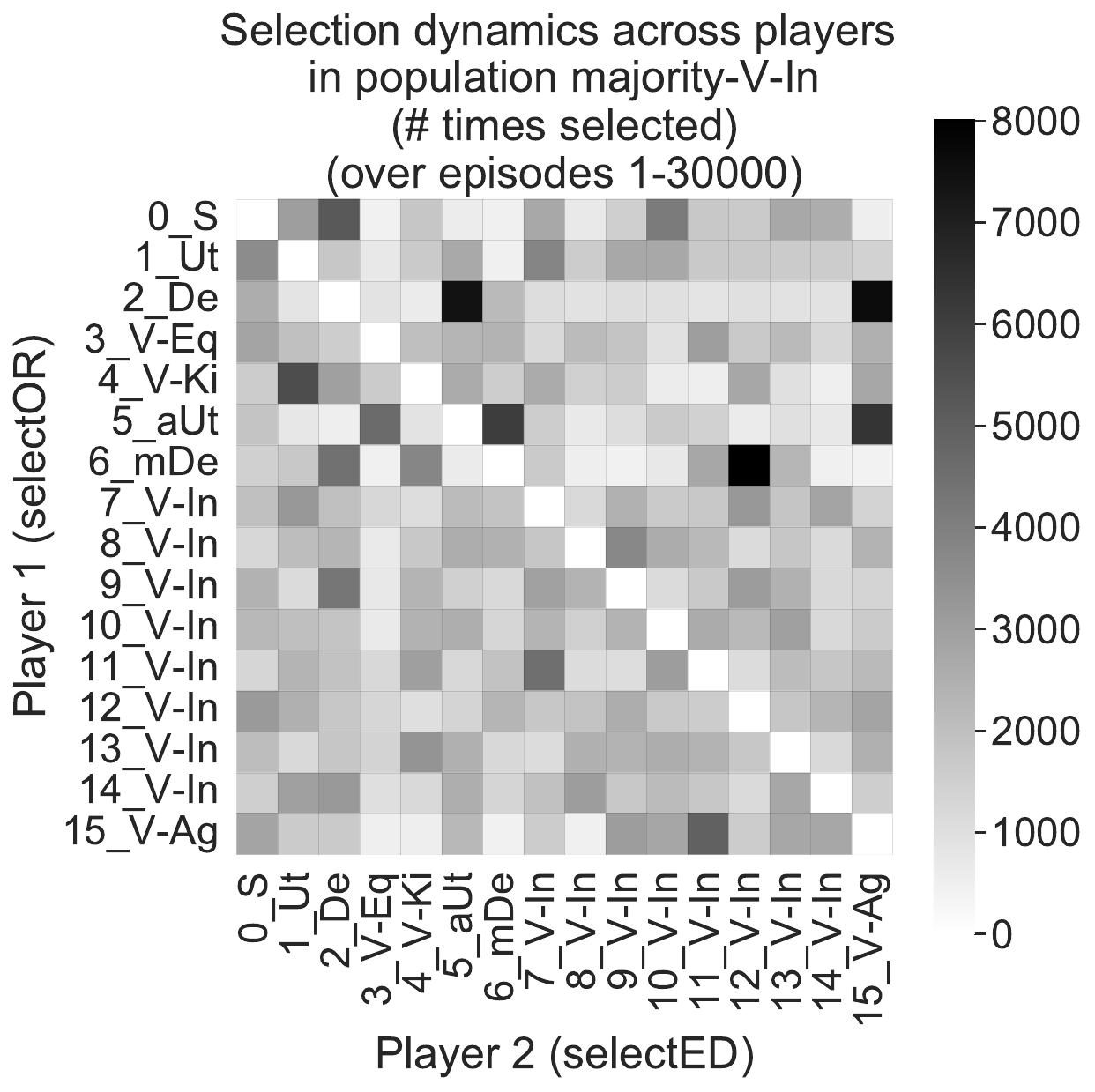}
        \\
        \includegraphics[height=0.45\linewidth]{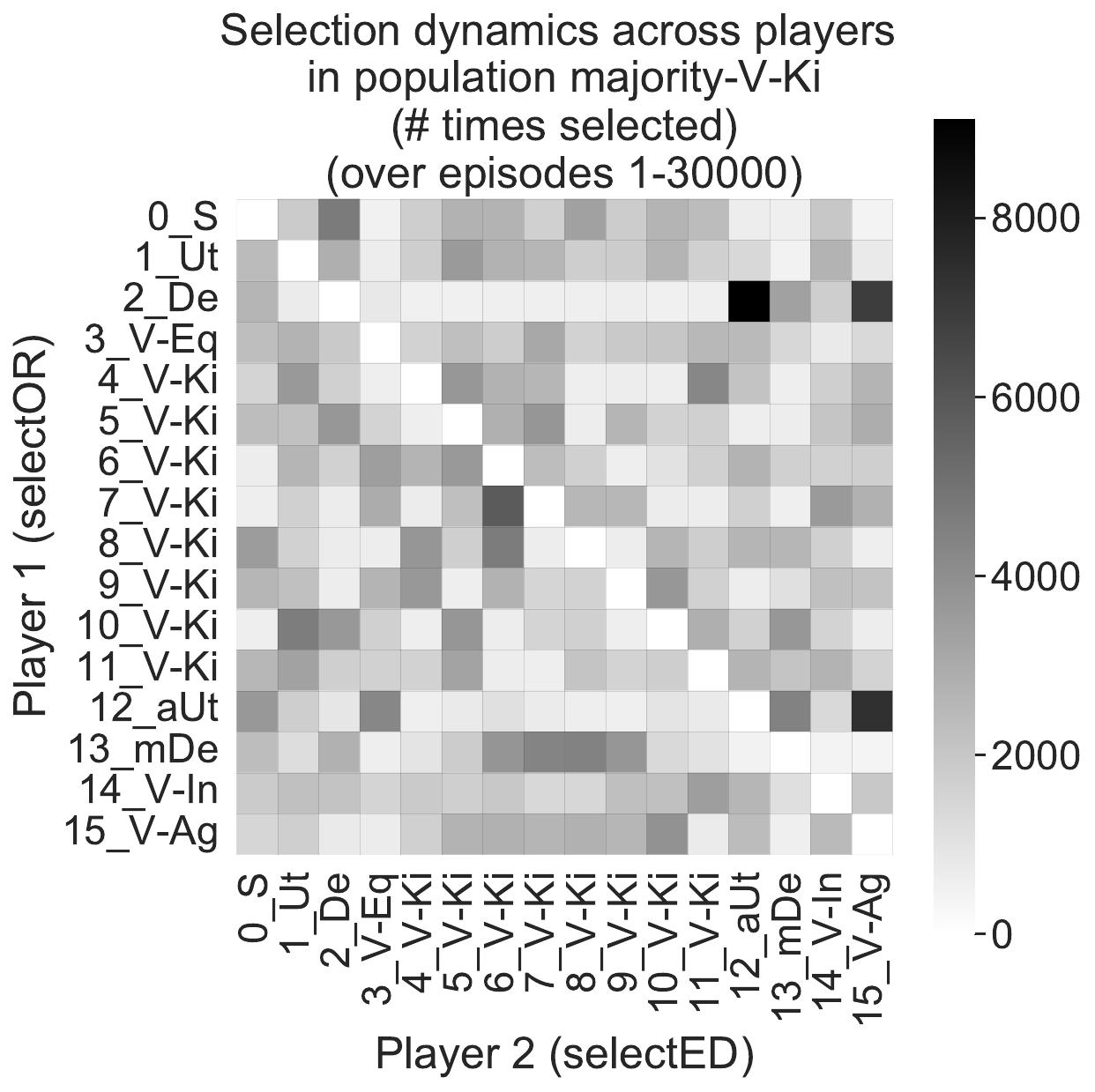}
        \includegraphics[height=0.45\linewidth]{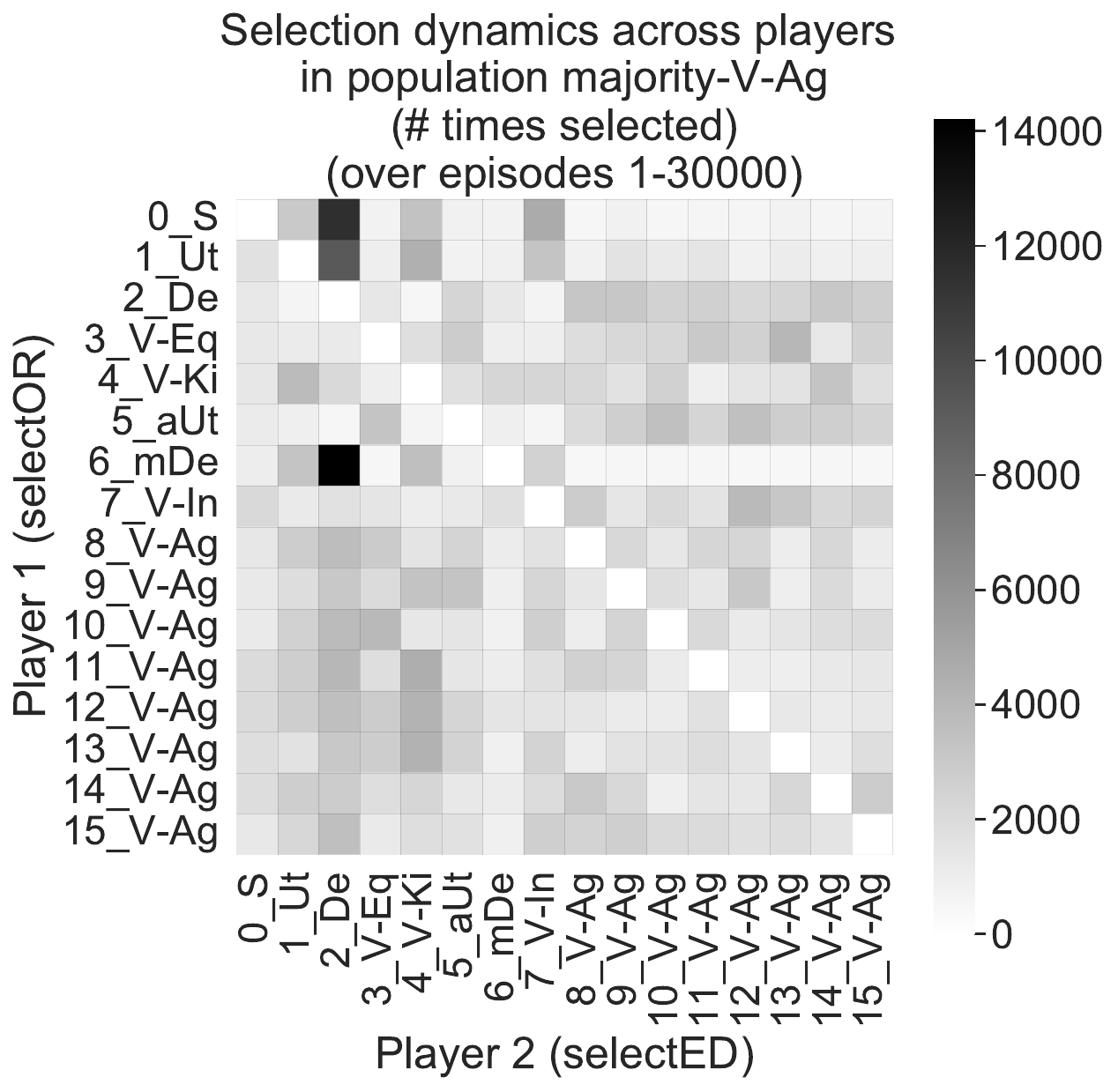}
    \end{center}
    Figure 8b: Selections made by individual players in populations \textit{majority-V-Eq}, \textit{majority-V-In}, \textit{majority-V-Ki} and \textit{majority-V-Ag}. The number of selections in every cell is summed over all 30000 episodes (average across 20 runs).
    \label{fig:heatmap_selections_byplayer}
\end{figure*}

\begin{figure*}[h]
    \begin{center}
        \includegraphics[width=0.39\linewidth]{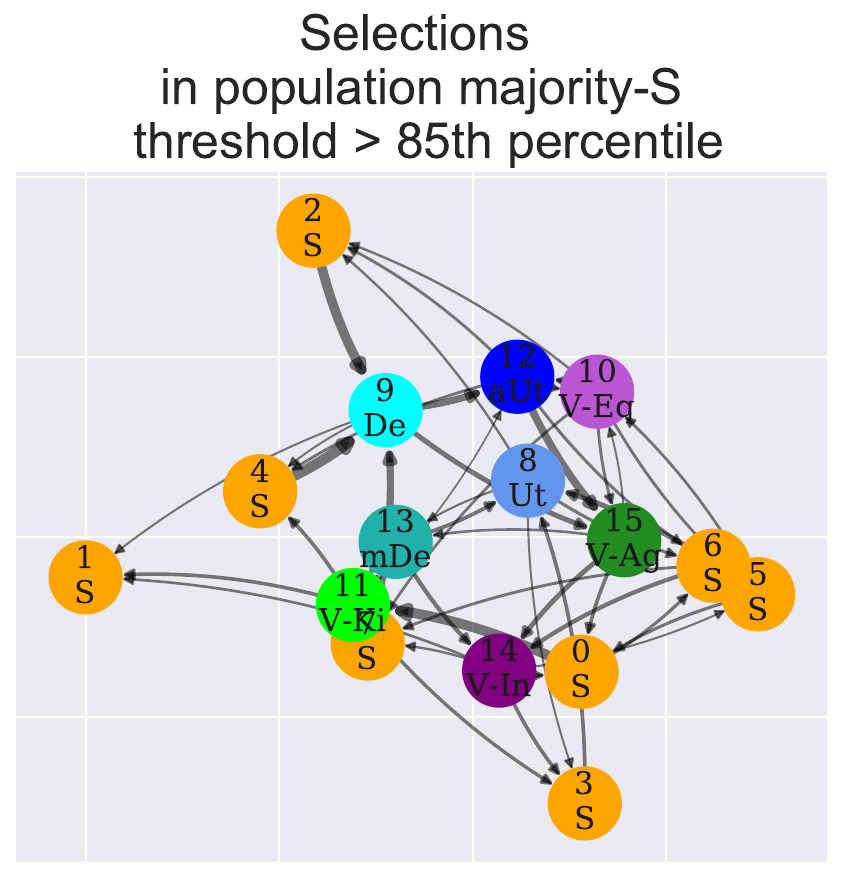}
        \\ 
        \includegraphics[width=0.39\linewidth]{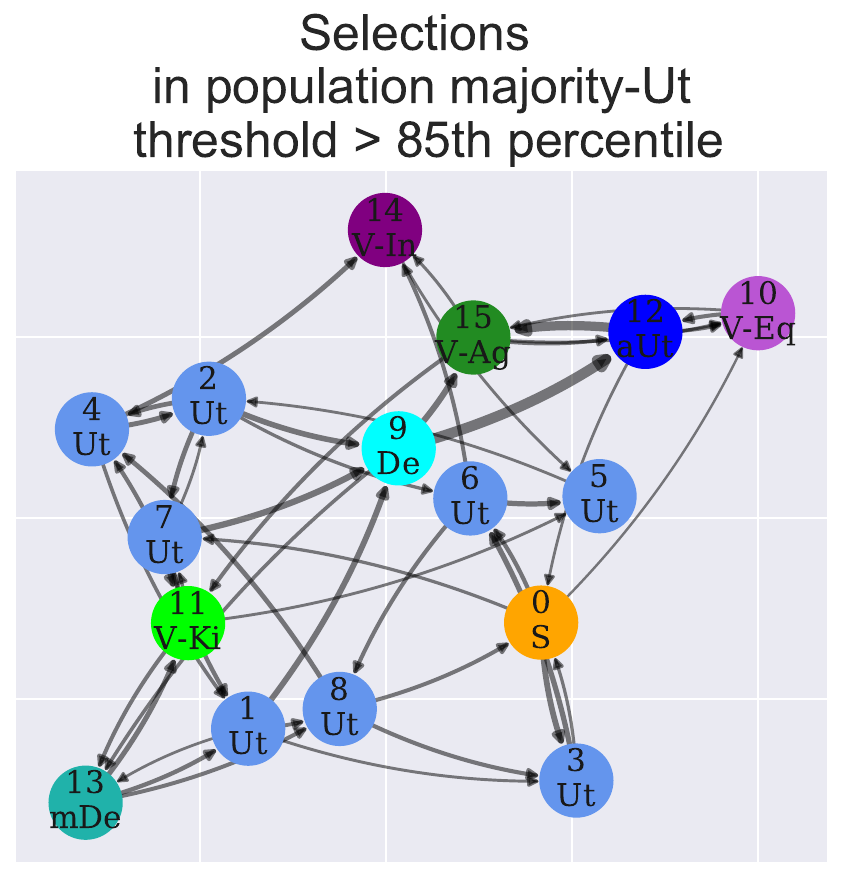}
        \includegraphics[width=0.39\linewidth]{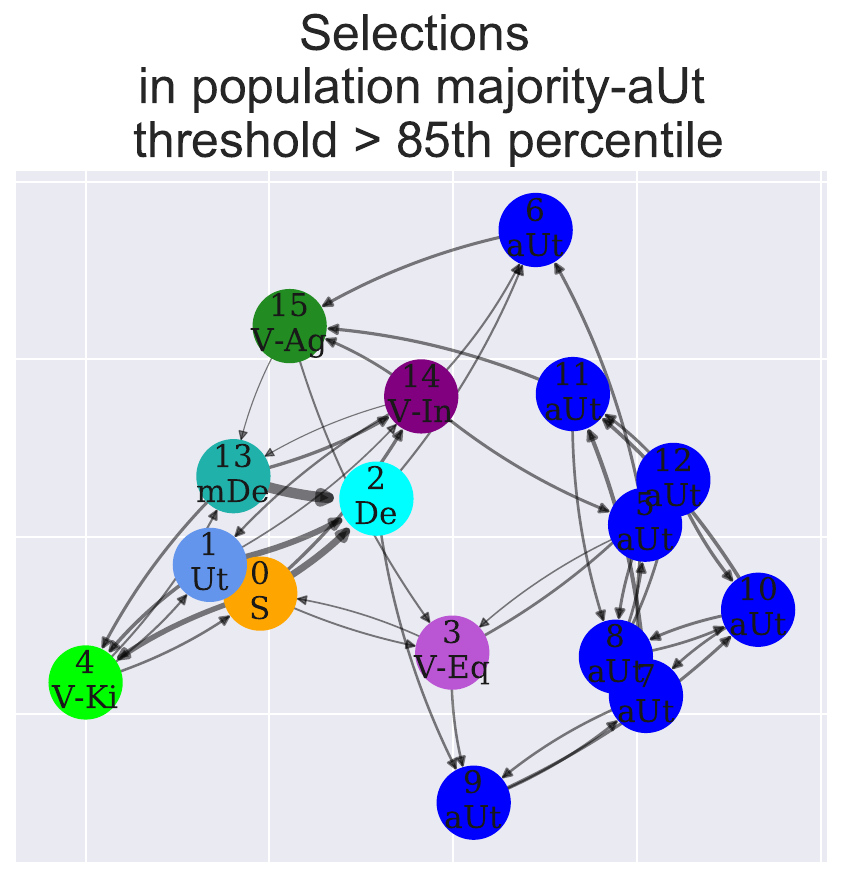}
        \\
        \includegraphics[width=0.39\linewidth]{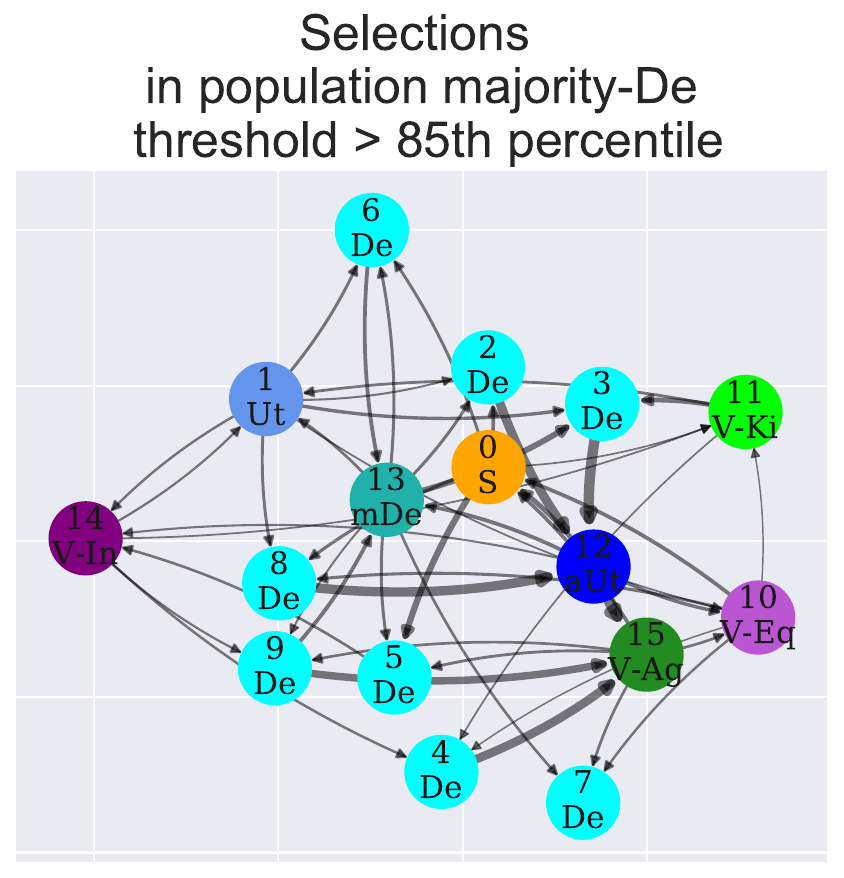}
        \includegraphics[width=0.39\linewidth]{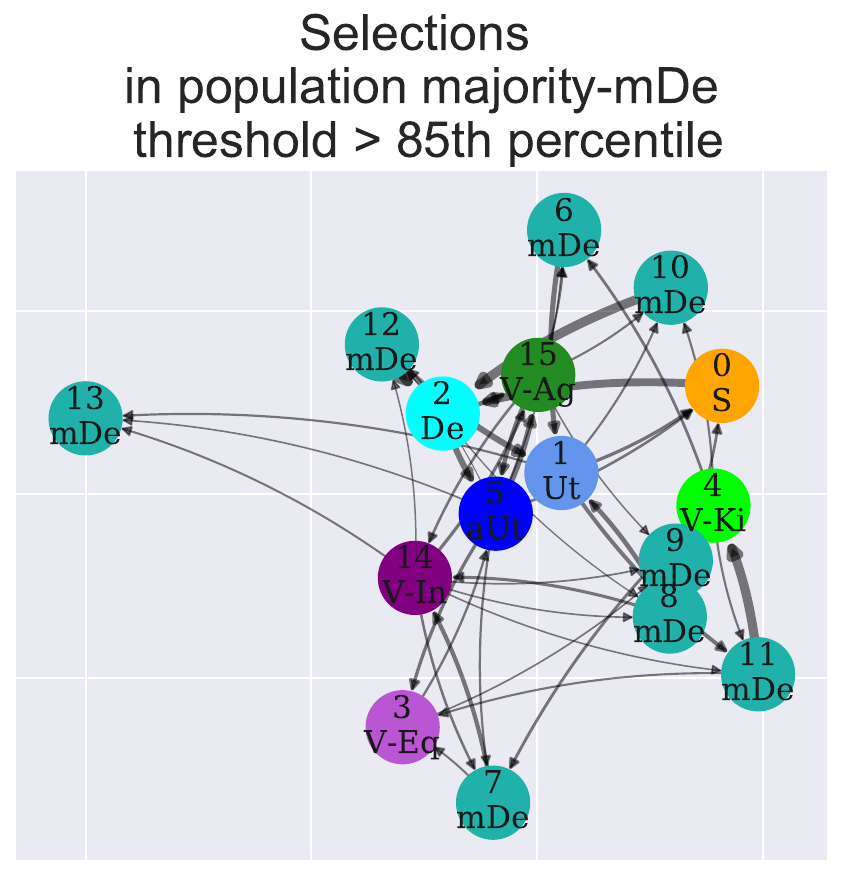}

    \end{center}
    Figure 9a: Selections made by individual players in populations \textit{majority-S, majority-Ut, majority-aUt, majority-De} and \textit{majority-mDe}. Edges are weighted by the number of selections made between the pair of players. We visualize the most common interactions, i.e., above the 85th percentile, over all 30000 episodes. More popular players are placed more centrally using a force-directed algorithm.
\end{figure*}

\begin{figure*}[h]
    \begin{center}
        \includegraphics[width=0.39\linewidth]{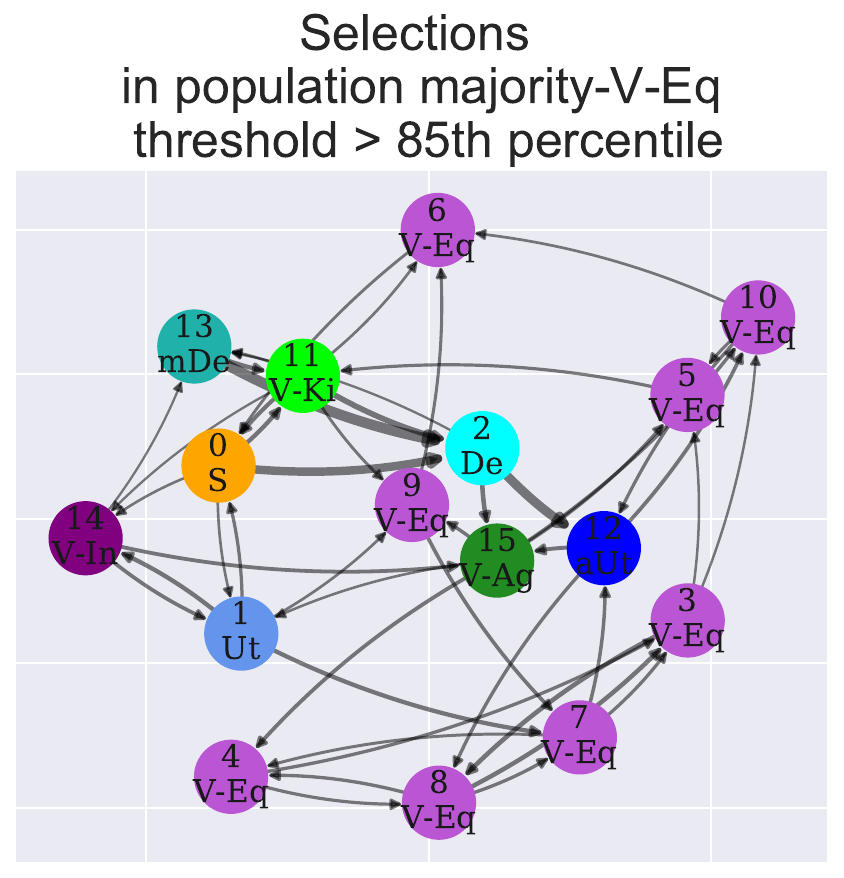}
        \includegraphics[width=0.39\linewidth]{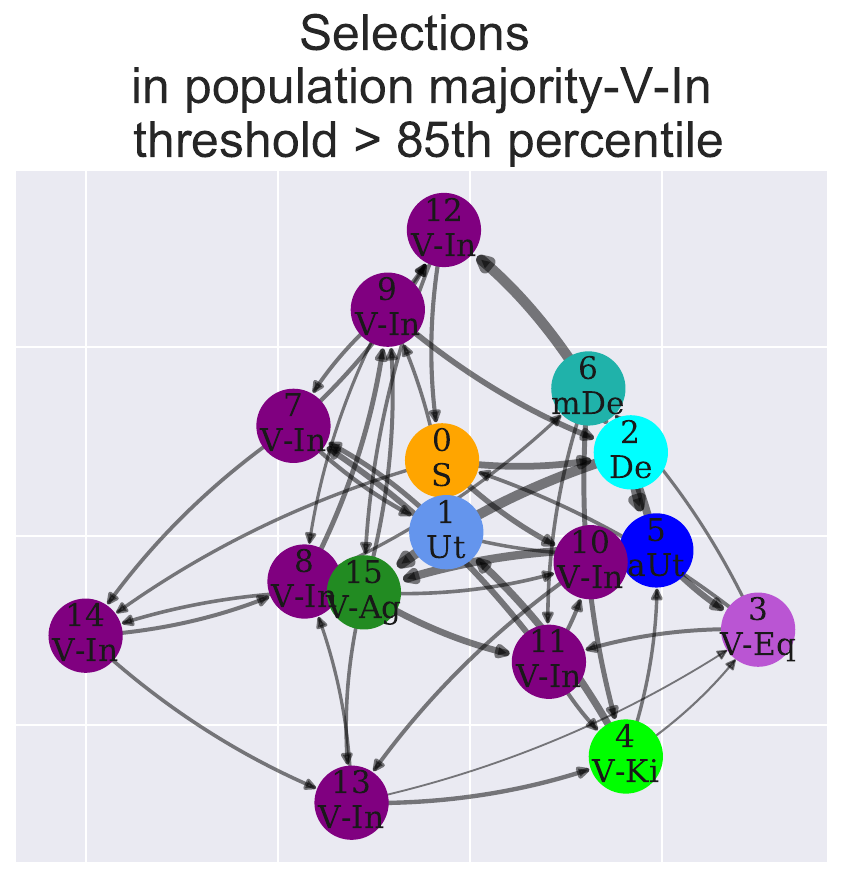}
        \\
        \includegraphics[width=0.39\linewidth]{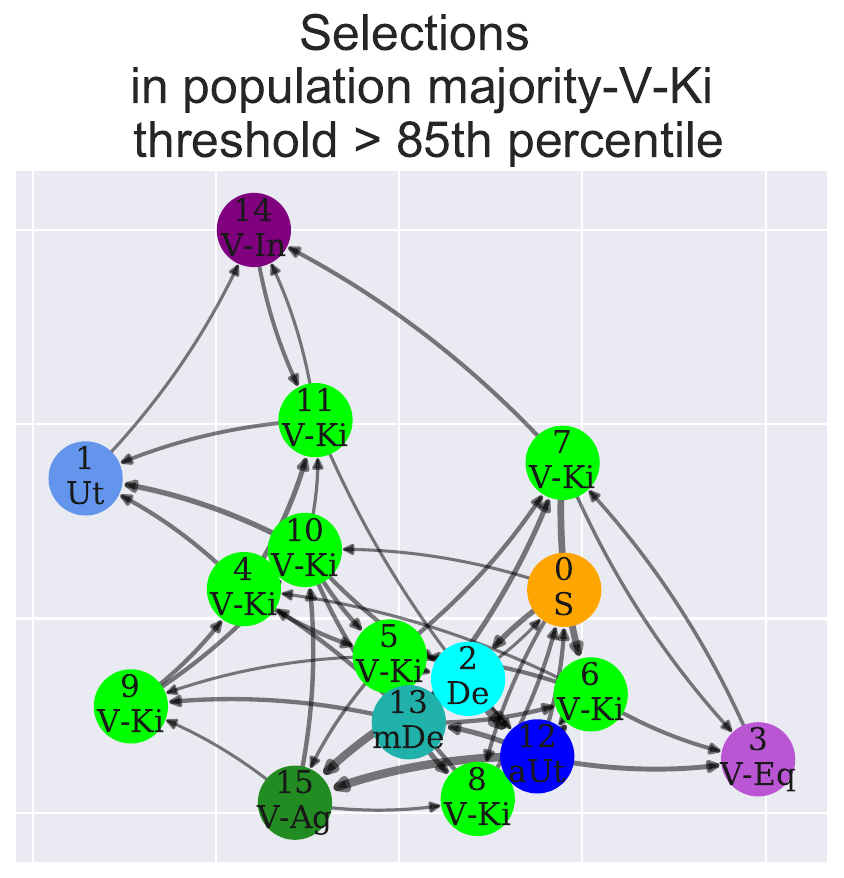}
        \includegraphics[width=0.39\linewidth]{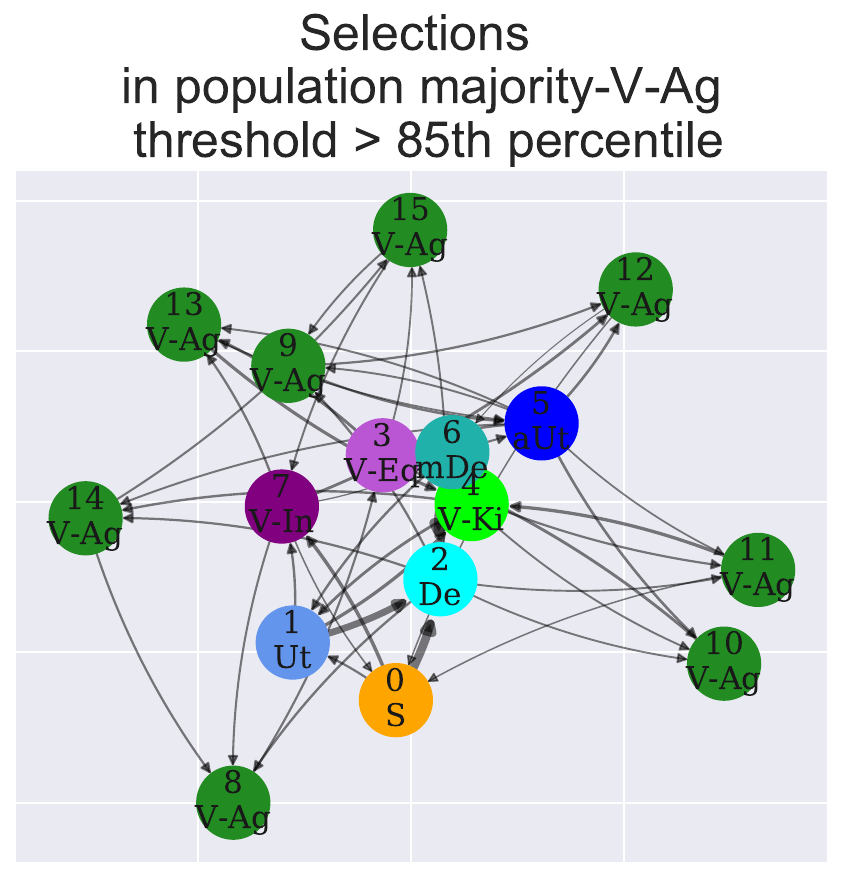}
    \end{center}
    Figure 9b: Selections made by individual players in populations \textit{majority-V-Eq}, \textit{majority-V-In}, \textit{majority-V-Ki} and \textit{majority-V-Ag}. Edges are weighted by the number of selections made between the pair of players. We visualize the most common interactions, i.e., above the 85th percentile, over all 30000 episodes. More popular players are placed more centrally using a force-directed algorithm.
    \label{fig:networks_bipartite}
\end{figure*}

\section{Partner Selection Algorithm}
\label{sec:appendix_algorithm}

The algorithm for partner selection ($sel$) and at the basis of the dilemma game ($dil$) with 2 separate Q-network models is presented below (see Algorithm \ref{alg:ps}). We use episodes, so that all players learn simultaneously, after each player had a single chance to select and play.
For clarity, each selector's state $s_{i,sel}^{t}$ relies on the same $environment\_state$ (i.e., simultaneous selections), and therefore every player $i$'s $s_{i,sel}^{t+1}$ in the $sel$ trajectory only differs from $s_{i,sel}^{t}$ by one entry: $a_{j,dil}^{t}$ at player $j$'s position. 
For $dil$, each player $i$ uses \textit{all} of their observed dilemma trajectories to update their network, by calculating an average loss across these trajectories.

We use the DQN algorithm for agents learning to select and to play the dilemma \citep{mnih2015human}, with a slight simplification: we refresh the experience buffer at the end of every episode (see Algorithm \ref{alg:ps}), and we do not freeze the network weights (experiments showed that freezing the network made no impact on the stability of the learning).

\begin{algorithm*}[h]
\small
\caption{Partner selection algorithm}\label{alg:ps}
\begin{algorithmic}[1]

 \REQUIRE population of players $P$ of size $N$, total episodes $T=30000$
\STATE set random initial $environment\_state$ (i.e., one random move for each player) 
\FORALL {players $i \in P$} 
    \STATE initialize selection memory $D_{i, sel}$ and dilemma memory $D_{i, dil}$ (capacity $256$ to allow experience replay)
    \STATE initialize value functions $Q_{i,sel}, Q_{i,dil}$ 
    \STATE set random initial selection \& dilemma states $(s_{i,sel}^{t=0}, s_{i,dil}^{t=0})$ 
\ENDFOR 

\STATE initialize iterations counter $t=0$

\FORALL {${episode}\gets 1, T$}
    \STATE Get current global $environment\_state$ from the environment 
    
    \STATE \textbf{PART 1: Partner Selection}    
    \FORALL{players $i \in P$} 
        \IF{${episode}=0$}  
        \STATE use random state stored in $s_{sel}^{t=0}$ on first episode
        \ELSE{} 
        \STATE $i$ observes the $environment\_state$ and removes itself to parse in their own $s_{i,sel}^{t}$  
        \ENDIF
    \STATE $t \gets t+1$
    \STATE using $s_{i,sel}^{t}$, $Q_{i,sel}$ and $\epsilon$-greedy policy, $i$ selects a partner $j$ (i.e., takes action $a_{i,sel}^t$) 
    \STATE store pair of agents $\{i,j\}$ in this episode's sub-population $P_{episode}$
    \ENDFOR \\

    \STATE \textbf{PART 2: Dilemma Game} 
    \FORALL {$\{i,j\}\in P_{episode}$} 
        \IF{$episode=0$} 
        \STATE use random dilemma state stored in $s_{dil}^{t=0}$ on first episode
        \ELSE{}
        \STATE $i$ and $j$ parse states $s_{i,dil}^t$, $s_{j,dil}^t$ from the $environment\_state$ \& the current opponent's index 
        
        \ENDIF 
        
        \STATE using $s_{i,dil}^{t}$, $Q_{i,dil}$ and $\epsilon$-greedy policy, $i$ plays a move $a_{i,dil}^t$
        \STATE using $s_{j,dil}^{t}$, $Q_{j,dil}$ and $\epsilon$-greedy policy, $j$ simultaneously plays a move $a_{j,dil}^t$
        \STATE $i$ and $j$ receive rewards $r_{i,dil}^t$ \& $r_{j,dil}^t$ respectively
        \STATE $i$ and $j$ observe $s_{dil}^{t+1}$ based on the current opponent's action $a_t$: $s_{i, dil}^{t+1} = a_{j, dil}^t; s_{j, dil}^{t+1} = a_{i, dil}^t$

        \STATE player $i$ copies the dilemma reward to the selection reward as well: $r_{i,sel}^t=r_{i,dil}^t$ 
    
        \STATE in player $i$'s $D_{i, dil}$ memory, store: $(s_{i, dil}^t, a_{i, dil}^t, r_{i, dil}^t, s_{i, dil}^{t+1})$ 
        \STATE in player $j$'s $D_{j, dil}$ memory, store: $(s_{j, dil}^t, a_{j, dil}^t, r_{j, dil}^t, s_{j, dil}^{t+1})$ 
    \ENDFOR\\

    \STATE \textbf{PART 3: Update $sel$ and $dil$ networks} 
    \FORALL{players $i\in P$}
        \STATE update the corresponding part of $environment\_state$ with the latest moves by $i$ ($a_{i, dil}^t$)
        \STATE player $i$ observes their $s_{i,sel}^{t+1}$ by copying the latest $environment\_state$ and removing self 
        \STATE in player $i$'s $D_{i, sel}$ memory, store: $(s_{i, sel}^t, a_{i, sel}^t, r_{i, sel}^t, s_{i, sel}^{t+1})$ 
        
        \STATE player $i$ samples one experience from $D_{i, sel}$: $(s_{i, sel}^t, a_{i, sel}^t, r_{i, sel}^t, s_{i, sel}^{t+1})$ \& calculates loss
        \STATE player $i$ samples all experiences from $D_{i, dil}$: $(s_{i, dil}^t, a_{i, dil}^t, r_{i, dil}^t, s_{i, dil}^{t+1})$ \& calculates average loss

        \STATE player $i$ updates $Q_{i, sel}$ network with respect to time step $t$ \& the calculated loss 
        \STATE player $i$ updates $Q_{i, dil}$ network with respect to time step $t$ \& the calculated loss
        \STATE refresh memory buffers $D_{i, sel}$, $D_{i, dil}$, $D_{j, dil}$
        
    \ENDFOR 

\ENDFOR 
\end{algorithmic}
\end{algorithm*}

\end{document}